\documentclass[manuscript,screen,nonacm]{acmart}
\usepackage{subcaption}
\usepackage{array}
\usepackage{tabularx, booktabs, xspace}

\newcommand{\cf}{cf.\ }
\newcommand{\eg}{e.g.,\ }
\newcommand{\etal}{et al.\xspace}
\newcommand{\ie}{i.e.,\ }

\usepackage[capitalize,noabbrev]{cleveref}

\newcommand\clearrow{\global\let\rowmac\relax}

\crefformat{section}{\S#2#1#3}
\crefformat{subsection}{\S#2#1#3}
\crefformat{subsubsection}{\S#2#1#3}

\newcommand{\para}[1]{{\vspace{.05in} \bf \noindent #1 }}
\newcommand{\parait}[1]{{\vspace{.05in} \em \noindent #1 }}

\AtBeginDocument{%
  }

\begin{document}

\title[Wording Matters]{Wording Matters: Understanding How GenAI Policy Rules Shape Teachers' and Students' Workflows and Perceptions}

\author{Sabiha Mehreen}
\email{sabiha-mehreen@uiowa.edu}
\affiliation{%
  \institution{University of Iowa}
  \city{Iowa City}
  \state{Iowa}
  \country{USA}
}

\author{Xiangzhe Yuan}
\affiliation{%
  \institution{University of Iowa}
  \city{Iowa City}
  \state{Iowa}
  \country{USA}}
\email{xiangzhe-yuan@uiowa.edu}

\author{Rishab Nithyanand}
\affiliation{%
  \institution{University of Iowa}
  \city{Iowa City}
  \state{Iowa}
  \country{USA}}
\email{rishab-nithyanand@uiowa.edu}

\author{Naimul Hoque}
\affiliation{%
  \institution{University of Iowa}
  \city{Iowa City}
  \state{Iowa}
  \country{USA}}
\email{naimul-hoque@uiowa.edu}

\renewcommand{\shortauthors}{Mehreen et al.}

\begin{abstract}
 The growing use of generative AI (GenAI) in higher education has led
institutions to develop policies governing its applications. However, we have little knowledge about how teachers and students adapt and adhere to the policies. In this paper, we analyzed GenAI policies from 101 U.S.\ universities, identifying nine recurring
rules for GenAI use, and interviewed 20 students and 20 faculty members about
the rules. We find that the wording of the rules can change how students perceive and act on the rules while introducing different challenges that teachers face in enforcing them in practice. For example, rules such as \textit{``Passing off AI-generated content as your own constitutes academic dishonesty''} and \textit{``Students are accountable for any inaccurate, biased, or unethical content''} both outline the acceptable use of AI, but the former can make students hide their use of AI out of fear of unwanted scrutiny, whereas the latter can encourage students to take responsibility and ownership of their use of AI and be more transparent. Students described uncertainty about acceptable use, fear of penalties and unfair treatment, and stigma that discouraged acknowledging AI use. Teachers took on approval workload, had to interpret acceptable use, and lacked reliable ways to verify compliance. We summarize these impacts in a policy-impact matrix linking each rule to its effects on teacher and student workflows and perception.
\end{abstract}

\keywords{GenAI, Attribution, Policy Rules, Taxonomy, Interviews}

\maketitle

\section{Introduction}\label{sec:intro}
Universities are developing policies and guidance to establish acceptable uses of generative AI (GenAI) in academic work.
These policies extend beyond blunt statements about whether students may use GenAI to more nuanced descriptions about what constitutes authorship, disclosure requirements, and responsibility for generated material~\cite{jin2025generative, wang2024generative}.
Applying these policies, however, requires judgment that the policies themselves do not settle.
For example, a requirement to submit one's own work leaves open questions about whether AI-assisted brainstorming is acceptable, while a disclosure requirement leaves open questions about what exactly must be reported and how it will be evaluated.
These questions must be resolved by students when completing their work, and by faculty when communicating expectations and assessing compliance.
Understanding how both groups approach these judgments is therefore necessary to connect institutional rules with academic practices.

Prior research has broadly characterized institutional approaches to GenAI \cite{wang2024generative} and documented challenges related to policy ambiguity \cite{tsao2025trajectories}, undisclosed assistance \cite{adnin_examining_2025, gonsalves2025addressing}, and the effort required to assess suspected violations \cite{lau2026barriers}.
These findings establish that applying GenAI policies can be challenging, but leave open how specific policy requirements contribute to those challenges.
For example, when students hesitate to disclose GenAI usage, is it because the requirement is unclear, the associated workload is overly burdensome, or they anticipate unfavorable treatment?
These distinctions are important because they suggest which changes to GenAI policies are likely to improve alignment between institutional rules and academic practices, and result to better educational outcomes for students.

\para{Our contributions.}
Our study combines an analysis of institutional GenAI policy documents with interviews of students and instructors to address this gap.
We use recurring rules extracted from the documents as the basis for our interview prompts, thus grounding stakeholder accounts in expectations communicated by universities.

Specifically, we address the following research questions:

\parait{RQ1. What recurring expectations govern students' GenAI use and how are they communicated? (\Cref{sec:policy_analysis})}
    To answer this question, we analyzed 694 policy documents from 101 US institutions using a combination of manual thematic analysis and AI-assisted annotation.
    We identified nine frequently recurring GenAI-related rules or expectations, each appearing in materials from at least 77 institutions, with different institutional sources emphasizing different expectations.
    This distribution of expectations across institutional sources suggests that understanding acceptable GenAI use may require consulting several different sources and determining how their requirements apply together.

\parait{RQ2. How do students and instructors interpret and apply these rules, and what are the implications for policy design? (\Cref{sec:interviews})}
    We used the nine rules as common prompts in interviews with 20 students and 20 instructors, then analyzed their accounts.
    Participants described differences in how they interpreted these rules, focusing on issues related to uncertainty and compliance burdens.
    Broadly, these findings reinforce that policies need to account for both what students are expected to do and what instructors can reasonably assess.
    Clear requirements should therefore be accompanied by practical ways to demonstrate compliance and explicit expectations about how that information will be evaluated.
    Otherwise, efforts to improve accountability may add work for both groups while leaving students uncertain about the consequences of reporting their use.

\noindent Based on these findings, we discuss approaches for improving the development and enforcement of GenAI policies (\Cref{sec:discussion}).

\section{Related Work} \label{sec:related}

\para{GenAI policies combine institutional expectations with instructor discretion.}
Generative AI (GenAI) offers new opportunities to support student learning and academic work, as well as concerns related to academic integrity and assessment \cite{lo2023impact,yusuf2024generative}. 
Policy analyses document how institutions address these concerns through guidance and  resources \cite{dabis2024ai,mcdonald2025generative,MOORHOUSE2023100151, an2025investigating, dai_why_2025}. 
For example, Jin \etal \cite{jin2025generative} examined GenAI usage policies from six global regions and identified a recurring emphasis on academic integrity, equity, and teaching and learning.
Other studies found that institutions set broad principles and left decisions related to the adherence of these principles to instructors \cite{wang2024generative, qian2026governing, lafrance2026governing}.
Notably, in 2024, Wang \etal \cite{wang2024generative} found that more than half of the Top 100 US universities delegated decisions about permitted use to instructors and their guidance was primarily addressed to faculty rather than students.
Our policy analysis (\Cref{sec:policy_analysis}) builds on these characterizations by asking which specific expectations recur and where they are communicated.
We analyze 694 documents sourced from 101 institutions' central offices, teaching centers, and disciplinary units, identifying nine recurring expectations and comparing their prevalence across different institutional sources.
Our findings confirm that widespread deference to instructors exists, but also shows that different expectations are communicated by different institutional sources. 
This suggests a potential challenge  related to information fragmentation for instructors (and students) to understand the rules governing GenAI use in the classroom.

\para{Operationalizing policies increases burden for instructors and students.}
Studies have shown that the deference to instructors results in different interpretations.
For example, Wu \cite{Wu2026Governing}, in a study of 35 syllabi at one university found wide variance in GenAI policies at the course-level.
Additionally, Tong \etal \cite{tong2025we} found variations in the linguistic style and framing of GenAI policies in syllabi.
In a study by Lau \etal \cite{lau2026barriers}, instructors described the challenges associated with this deference under non-prescriptive guidance as ``policy crosswinds'' and indicated that students were confused by the different expectations in each class.
Further complicating matters for instructors are challenges related to the enforcement of their expectations which is limited by the poor accuracy \cite{weberwulff2023testing, sadasivan2023can} and known biases \cite{liang_gpt_2023} of AI-text detection tools.
Corbin \etal \cite{corbin2025talk} characterize the reliance on unenforceable rules as having an ``enforcement illusion''.
Research has also focused on student perspectives.
Gonsalves \etal \cite{gonsalves2025addressing} found that even when reporting was required, many students did not report their GenAI use, citing fear of consequences, ambiguity in policies, and inconsistent enforcement.
Uncertainty about instructor judgment of reported use also strains the trust between them \cite{luo2025does}.
Prior work also showed the effects of placing the burden of proof of compliance on students.
Students contesting accusations of non-compliance described the stress associated with collecting draft versions and screen recordings to prove their innocence \cite{gorichanaz2023accused, nytimes_headache}.
Teaching guidance now encourages students to document their process before accusations arise \cite{stover_proving_2026} and courses are increasingly formalizing this record keeping through reporting frameworks and tools that capture the writing process \cite{perez-perez_transparency_2026, weaver2024artificial, kim2026attribution, siddiqui_draftmarks_2025}.
However, how much documentations is reasonable to maintain remains unclear \cite{Hosseini2025}.
Our interview findings (\Cref{sec:interviews}) uncover a related tension: brief disclosures are manageable to students yet insufficient for instructors to assess the nature of the AI assistance.

\para{Policy framing and ambiguity influences their interpretation and consequences.}
Prior research has examined how policy language frames originality \cite{luo2024critical}, communicates expectations \cite{moore2024communicating}, and guide policy development \cite{chan2023comprehensive}.
Communication studies examine relationships between instructor messaging and student perceptions \cite{kaufmann2026trend, vallade2025generative}, including responses to academic integrity concerns \cite{petricini2025ai}, and how framing predicts perceptions on AI use.
Jiang \etal \cite{jiang_exploring_2025} found that students and faculty considered institutional guidance ambiguous.
Further, Tsao \cite{tsao2025trajectories} identified that this ambiguity led to diverse interpretations, anxieties, and inequities in stakeholder experiences.
One consequence of this ambiguity and choices in policy framing relates to students' development of strategies and justifications for concealing GenAI use and teachers' responses to non-disclosure \cite{adnin_examining_2025}.
These consequences were also highlighted by Qu \etal \cite{qu2025generative} who described students viewed AI assistants as personal tools rather than contributors needing acknowledgments.
Together, these findings show that non-compliance can reflect understandings of GenAI policies, as well as concerns and uncertainty about how compliance will be perceived or assessed.
Our study builds on these explanations by connecting stakeholder concerns with commonly stated expectations.

\section{Policy Analysis} \label{sec:policy_analysis}
Our analysis asks: {\em  what recurring expectations govern students' GenAI use and how are they communicated?}
These expectations may appear across institutional policies, academic-unit-specific guidance, and similar expectations may be expressed in different terms.
To answer this question, we used a scaffolded approach that combined manual interpretation with AI agent-assisted data collection and annotation.
We first examined institutional websites manually to understand the range of available guidance and to formalize inclusion and exclusion criteria for relevant policy pages. 
These criteria guided an automated AI agent in collecting a corpus of GenAI-related policy documents. 
We then manually annotated a subset of the collected policies to identify recurring themes, which we refer to as ``{\em rules}'', related to the use of GenAI in the academic work.
We translated these rules into a codebook and supplied it, together with manually annotated examples, to an AI agent to guide annotation of the complete corpus of GenAI policy documents.
Finally, we assessed the quality of the agent's annotations through a separate manual review.
This approach allowed us to examine recurring rules across a large corpus of GenAI policy documents while grounding their interpretation in manual analysis, providing a basis for the subsequent student and instructor interviews.

\subsection{Identifying and collecting policy sources}\label{sec:policy_analysis:sources}

We used the institutional list compiled by Wang \etal \cite{wang2024generative} as the sampling frame for our analysis. After removing incomplete items, the supplied list contained 101 unique institutions which belonged to the 2024 US News Best National University Rankings.
These institutions offered educational programs across a range of disciplines and degree levels, providing a useful context for examining how GenAI usage expectations are communicated across academic units.
However, the focus on highly ranked universities limits the representativeness of our sample, as their resources and policy development practices may differ from other educational institutions.
For each institution in this list, we searched for publicly available GenAI usage guidance using GenAI-related keywords (\eg ``generative AI'', ``AI'', ``GenAI''), combined with the institution name and terms related to policy guidance (\eg ``policy'', ``guidance'', ``syllabus'', ``usage''), students, faculty, teaching, and learning.
We manually examined the results of this search to understand where institutions communicated expectations of GenAI use and which materials were relevant to our study. 
This initial review informed the inclusion and exclusion criteria used in our subsequent AI agent-assisted search. 
Specifically, we included institutionally issued documents that addressed GenAI, either wholly or in part, and established boundaries for students' use through examples or explicit rules.
We excluded general teaching or pedagogical policies, policies that focused on instructors' use of GenAI rather than students' use, and materials outside the institutional source scope, including individual instructors' course syllabi.
We provided these criteria to an AI agent and instructed it to search each university's official domains for policy documents that satisfied the supplied criteria, focusing on their central offices and academic units in sciences, engineering, humanities, social science, law, and medicine.
Each matching document was retrieved and saved by the agent. 
The complete prompt used for this task is illustrated in \Cref{fig:policy_analysis:sources:prompt}.

\para{Collection results.} 
Our final corpus included 694 policy and guidance documents across 101 unique institutions. \Cref{fig:policy_analysis:sources:breakdown-by-unit,fig:policy_analysis:sources:policies-per-institution-cdf} in \Cref{app:policies} illustrate their breakdown across institutional sources.
Teaching centers contributed the largest number of GenAI policy documents (153 documents across 71 different institutions), followed by Provost offices (77 across 51 institutions), and medical/health-profession units (72 across 40 institutions).
Among the 101 institutions in our corpus, the median number of unique GenAI policies per institution was 6 and the median number of institutional GenAI policy sources was 4.
This distribution suggests that students and faculty may need to consult guidance from a range of sources including university administrative offices, teaching centers, governance units, and their own academic units to understand expectations for GenAI use.

\subsection{Manual annotation and rule extraction} \label{sec:policy_analysis:rules}
From our corpus, we selected 150 unique policies from 101 institutions for manual annotation using the grounded theory approach.
At first, we randomly selected 20 policies from 10 random institutions. 
Two researchers independently examined these policies to identify passages that expressed expectations about students' GenAI use.
The analysis focused on what a passage asked students or faculty to do, while preserving the context in which that expectation applied, including whether it appeared as a binding policy or simply as guidance. After the independent coding, the two researchers met to discuss and resolve disagreements. This resulted in the initial codebook, which was discussed with the full research team too. The rest of the policies from the manual coding set were divided equally between the two researchers.  During the coding process, they periodically met to discuss new observations and dimensions and updated the codebook accordingly. Boundary cases were discussed with the full research team for resolving confusions.

\para{Manual analysis results.}
Our manual analysis identified nine recurring rules governing students' use of GenAI and produced 432 manually annotated examples for the subsequent AI-assisted annotation. The nine rules are described in \Cref{tab:policy_analysis:rules}.
These rules each concern different obligations.
For example, requiring submitted work to reflect a student's own understanding (R2) establishes an expectation of authorship, while framing the presentation of AI-generated work as one's own as academic dishonesty (R4) establishes a higher expectation and makes a connection with misconduct.
Similarly, requiring students to disclose the use of AI (R5) does not, by itself, indicate whether undisclosed use constitutes a violation (R9).
The accountability rules (R6 and R7) distinguish responsibility for evaluating the generated outputs from restrictions on the information entered into the tool.

\begin{table*}[t]
\centering
\small
\setlength{\tabcolsep}{4pt}
\renewcommand{\arraystretch}{1}

\caption{The nine rules derived from the policy corpus, with the number of
excerpts and institutions coded to each rule, and one example excerpt.
An excerpt can be coded with more than one rule.}
\label{tab:policy_analysis:rules}

\begin{tabularx}{\textwidth}{@{}
    >{\arraybackslash}p{0.34\textwidth}
    >{\arraybackslash}X
    >{\centering\arraybackslash}c
@{}}
\toprule
\textbf{Rule} & \textbf{Example Excerpt} & \textbf{\# Excerpts} \\
{} & {} & {\bf (\# Institutions) } \\
\midrule

\textbf{R1 (Instructor Approval)}: Check with your instructor before using AI tools &
``Students must obtain permission from me before using AI composition software (like ChatGPT) for any assignments in this course.'' &
{2,508 (101)} \\
\addlinespace[2pt]

\textbf{R2 (Sole Ownership)}: All work submitted must be your own &
``All work submitted in this course must be your own, completed in accordance with the University's academic regulations.'' &
{831 (93)} \\
\addlinespace[2pt]

\textbf{R3 (Content Prohibition)}: Refrain from using AI tools to generate any content &
``You may not engage in unauthorized collaboration or make use of ChatGPT or other AI composition software.'' &
{990 (91)} \\
\addlinespace[2pt]

\textbf{R4 (Misrepresented Authorship)}: Passing off AI-generated content as your own constitutes academic dishonesty &
``Representing work that you did not produce as your own, including work generated or materially modified by AI, constitutes academic dishonesty.'' &
{329 (77)} \\
\addlinespace[2pt]

\textbf{R5 (AI Disclosure)}: Disclose when a work product was created wholly or partially using an AI tool &
``You may submit work that includes AI-generated content, but you must clearly indicate what was AI-generated and what was your own work.'' &
{1,841 (96)} \\
\addlinespace[2pt]

\textbf{R6 (Output Accountability)}: Student is accountable for any inaccurate, biased, or unethical AI-generated content &
``Because AI generated content is not necessarily accurate or appropriate, it is each student's responsibility to assess the validity and applicability of any generative AI output that is submitted.'' &
{1,420 (98)} \\
\addlinespace[2pt]

\textbf{R7 (Data Confidentiality)}: Never share sensitive, personal, or confidential information with AI systems &
``Any information you share with an AI tool is no longer private and may be used in ways you can't control.'' &
{1,344 (92)} \\
\addlinespace[2pt]

\textbf{R8 (Specified AI Uses)}: AI tools may be used for specific cases only &
``Generative AI may be used as a brainstorming tool to aid in information collection and organization.'' &
{1,121 (91)} \\
\addlinespace[2pt]

\textbf{R9 (Academic Penalties)}: Unacceptable use of AI will result in academic consequences including grade penalties &
``Engaging in unacceptable use of AI tools will result in academic consequences, which may include grade penalties, academic warnings, or other actions.'' &
{827 (87)} \\

\bottomrule
\end{tabularx}
\end{table*}

\subsection{Model-assisted annotation of the complete corpus} \label{sec:policy_analysis:annotation}
We used an AI-agent workflow to apply our manually developed nine-rule coding scheme to the complete corpus of 694 GenAI policy documents.
The workflow utilized five different agents (three annotation agents, one verification agent, and one adjudication agent), each driven by OpenAI's GPT6 Astra on `High' reasoning effort.
The three annotation agents were tasked with identifying and extracting all passages in the corpus that expressed any of the nine rules, using the rule definitions for each rule and all 432 manually identified examples as references.
A single excerpt could receive multiple labels when it expressed different obligations. For example, an excerpt could simultaneously grant permission to use GenAI for a specific activity and impose GenAI disclosure use (R8 and R5).
At the end of this stage, each annotation agent produced a collection of text excerpts that had specific rules associated with them, along with rationale for their relevance.
These were provided to a verification agent which assessed the validity (\ie if the produced output was a valid excerpt and rule) and quality (\ie does the output excerpt reasonably satisfy the inclusion/exclusion criteria for the output rule) of the match between the excerpt, assigned rule, and provided rationale.
If either verification test failed, the annotation agent was required to redo the annotation task and provided rationale for rejection by the verification agent.
Finally, when all annotation agents passed the verification tests, their outputs were provided to an adjudicator agent which resolved any recorded differences between the annotation agents' outputs by examining each rationale for inclusion/exclusion.
The prompts for each agent in this workflow are shown in \Cref{fig:policy_analysis:annotation:prompts} .

\para{Annotation validation.}
The agentic workflow identified 11,017 unique text excerpts with 11,211 rule assignments across 674 of the 694 policy documents in our corpus. The remaining 20 documents yielded no excerpts that matched our coding scheme.
Among all extracted excerpts, 3,480 (31.6\%) contained disagreements between the three independent annotator agents and required a final determination by the adjudicator agent.
To validate the output of our workflow, one researcher manually evaluated a separate sample of the annotations.
The review interface displayed each sampled excerpt within the context of the complete policy, together with the final accepted rules and rationale supplied by the adjudicator agent.
The researcher could then indicate whether the assignment was correct or incorrect, and indicate whether the excerpt was over-labeled (\ie assigned a rule that did not apply) or under-labeled (\ie missing a rule which did apply).
The sampling procedure prioritized balanced coverage across all 18 institutional source categories and 9 rules rather than random sampling. This was done to assess the correctness of annotations made in different policy contexts. In contrast, a random sample would have heavily biased validation towards descriptions of the `(R1) Instructor approval' rule in policies from Teaching Centers, which was the most frequently occurring (rule, institutional source)-pair. 
The researcher assessed 320 excerpts with 342 assigned rules from 186 documents, covering all institutional sources and all rules.
In total, the researcher confirmed 331 of the 342 assigned labels, identified 11 cases of over-labeling, and identified 3 cases of under-labeling.
The model performed uniformly well across all contexts, yielding a total F1-score of 97.9\% and per-rule F1-scores ranging from 96\% (R3: Prohibition) to 100\% (R1: Instructor approval).
Although our validation approach provides evidence of high-quality annotations from our agentic workflow, there is one important limitation that emerges due to the scale of the task: the validation evaluates the correctness of the workflow-identified (excerpt, rule)-pairs, rather than completeness.
Put differently, we did not evaluate whether the workflow missed excerpts that were relevant to specific rules. 
Doing so would have required manual annotation of a statistically meaningful number of complete policy documents, a prohibitively expensive task.
We expect, however, that our multi-agent workflow (with three independent annotators, one verifier, and one adjudicator) mitigates the impact of these errors.

\begin{figure}[t]
    \centering
    \begin{subfigure}{0.45\textwidth}
        \centering
        \includegraphics[width=\linewidth]{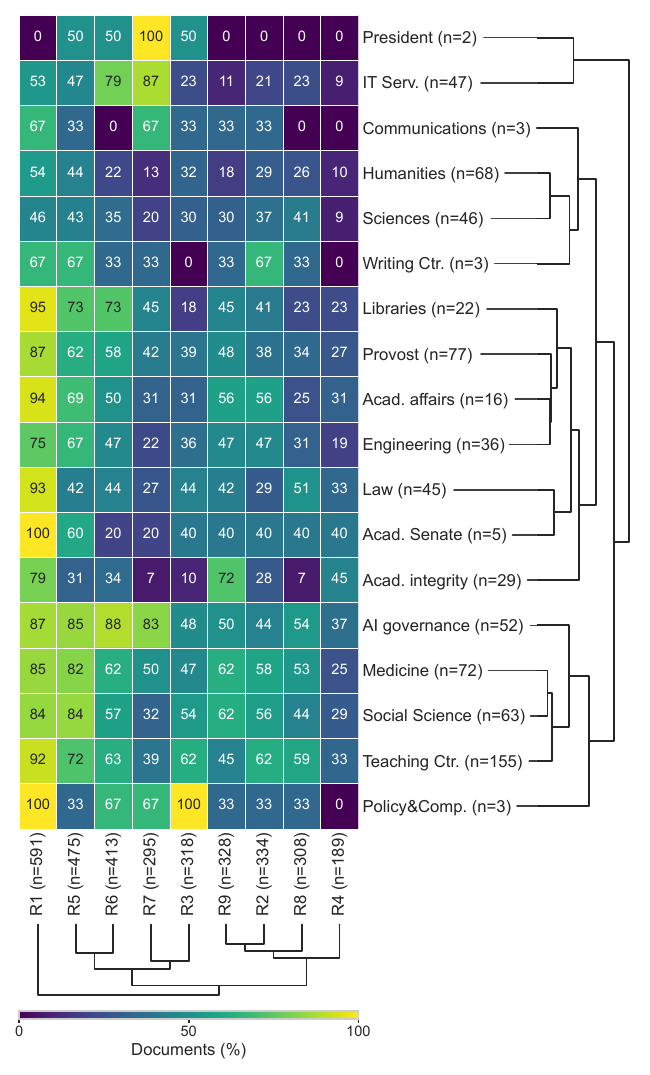}
        \caption{\%age of policy documents that refer to each rule.}
        \label{fig:policy_analysis:annotations:documents}
    \end{subfigure}
    \begin{subfigure}{0.45\textwidth}
        \centering
        \includegraphics[width=\linewidth]
        {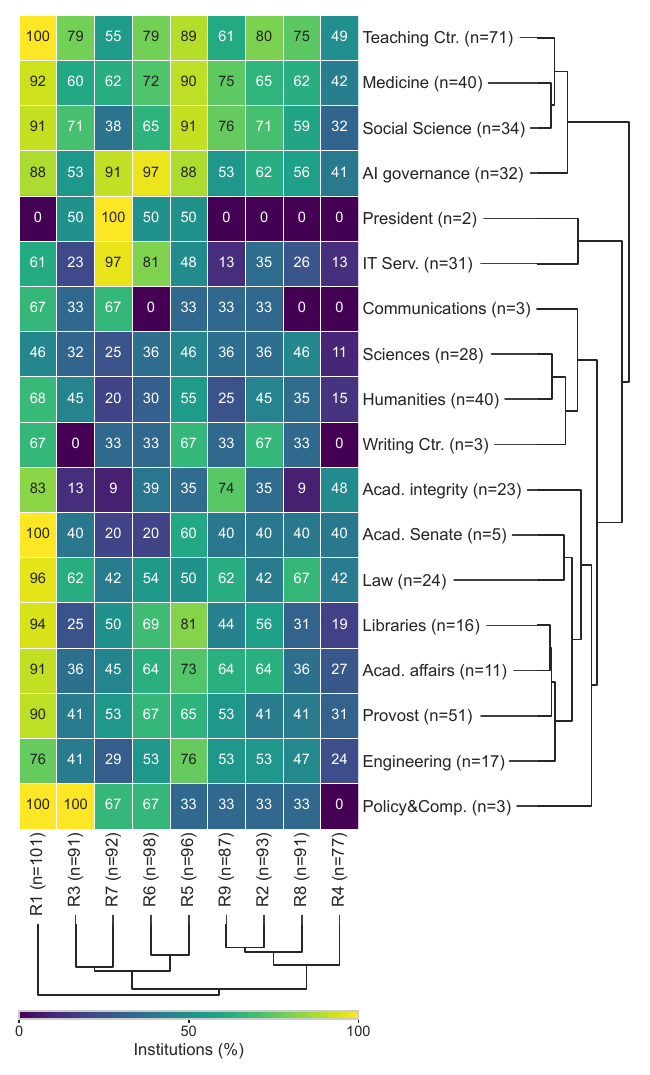}
        \caption{\%age of institutions with policies that refer to each rule.}
        \label{fig:policy_analysis:annotations:institutions}
    \end{subfigure}
    \caption{Prevalence of rules, broken down by institutional sources. Cells indicate the fraction of (a) unique documents or (b) institutions associated with the corresponding source that contain at least one reference to the corresponding rule. Each axis is clustered by the similarity of their profiles and their order varies between (a) and (b).}
    \label{fig:policy_analysis:annotations}
\end{figure}

\subsection{Annotation results } \label{sec:policy_analysis:results}
\Cref{tab:policy_analysis:rules} shows the number of text excerpts associated with each rule and the number of institutions with at least one policy document containing a relevant excerpt.
These aggregates indicate that the nine rules occurred widely across institutions and policy documents.
However, their broad presence does not establish that students and faculty encounter them together or understand how they apply to specific activities.
To better understand where rules are communicated, we analyzed their prevalence at the document- and institution-level broken down by their institutions sources.
For each institutional source, \Cref{fig:policy_analysis:annotations:documents} shows the percentage of documents containing each rule, while \Cref{fig:policy_analysis:annotations:institutions} shows the percentage of institutions with at least one document containing each rule.
We make several key observations. 

\parait{Deference to instructors is a common organizing principle in GenAI policies.} 
Statements making references to instructor authority or approval (R1) appeared in at least one policy document across all institutions, and were common across sources.
Their high prevalence suggests that institutional governance around AI often operates by assigning instructors responsibility for translating broad expectations into course-specific requirements.

\parait{Institutional sources emphasize different aspects of acceptable use.}
Guidance from IT services most frequently expressed statements related to data confidentiality (R7), appearing in 96.8\% ($n$=31) of the institutions represented in this category.
Academic integrity-related units showed a different emphasis: guidance from 73.9\% ($n$=23) of the represented institutions made statements about academic penalties (R9).
One explanation is that institutional offices approach GenAI through their existing responsibilities, producing guidance that is appropriate for each office's remit but requires stakeholders to combine requirements across multiple sources.

\parait{Disclosure requirements do not always carry misconduct framing.}
Disclosure and attribution requirements (R5) were more widely identified than statements characterizing misrepresented AI authorship as dishonesty (R4).
These differences reinforce the distinction between asking students to report AI assistance and explicitly connecting its misrepresentation to academic dishonesty.
This raises a question about how students interpret disclosure requirements: whether they understand disclosure as a standard work practice or associate it with the possibility of academic consequences.
It also simultaneously raises questions for faculty about how to communicate, evaluate, and enforce disclosure expectations.

\para{Takeaway.}
Broadly, our observations suggest that expectations surrounding the use of GenAI are widely articulated and commonly distributed across different institutional sources.
This creates a potential coordination problem: students and instructors must connect these distributed expectations that may each be communicated separately.
The resulting challenge, therefore, is not simply knowing whether GenAI is permitted, but understanding the conditions under which its use is acceptable and what demonstrates compliance.
This gap between identifying a rule and understanding how to interpret, operationalize (for faculty), or adhere to (for students) it motivates our interview study (\cf \Cref{sec:interviews}).
Our corpus and associated annotations are available \href{https://osf.io/23z4n/overview?view_only=6f8ed1d0e2864953a3312dfe65d14332}{here}.

\section{Interview Studies}\label{sec:interviews}
We conducted semi-structured interviews with 20 teachers and 20 students to examine the practical implications of the nine rules from Section \ref{sec:policy_analysis} in practice. The study was approved by our institution's Institutional Review Board (IRB).

\subsection{Participants}
We recruited teachers through LinkedIn posts, direct emails sent to publicly available institutional addresses, and snowball sampling. The final teacher sample comprised 20 participants (T01--T20) from universities across the United States who taught undergraduate and/or graduate courses and had GenAI policies for their respective courses. Their academic positions ranged from adjunct and teaching-track instructors to assistant and associate professors. Participants represented disciplines including computer science, sociology, writing and the humanities, engineering, physics, journalism, business, public policy, and biological sciences. We recruited undergraduate and graduate students through mass emails to several university listservs. Interested students signed up using a Qualtrics interest form. Of the 179 responses received, we purposively selected students to ensure variation in degree level, academic discipline, and self-reported GenAI use. We scheduled interviews with 20 students (S01--S20), all of whom completed their interviews. The student sample comprised 16 undergraduate and 4 graduate students from disciplines including health and life sciences, business, engineering and computing, education, and the humanities.  Additional details about participants' gender and academic backgrounds are provided in Table \ref{tab:Student_participants} and Table \ref{tab:teachers} in the Appendix.

\subsection{Interview Protocol and Analysis}
\label{sec:protocol}

We developed separate interview protocols for students and instructors around the nine GenAI attribution rules identified in our policy analysis. We refined both protocols through pilot interviews with four students and two instructors. During the interviews, we first presented the nine rules as patterns found in university policies. Next, we asked participants to select three or four rules they found relevant, confusing, difficult to follow or enforce, or connected to their experiences. This rule-selection activity served three purposes: 1) identifying which rules participants found most relevant or challenging, 2) eliciting their interpretations and experiences with specific rules, and 3) comparing how students and instructors understood and approached the same rules. For each selected rule, students described how they interpreted its requirements, whether they considered the rule reasonable, and whether they followed it in their own GenAI practices. Instructors discussed whether they used the rule in their courses, what challenges they had encountered or anticipated, and what other instructors should consider before adopting it. We used follow-up questions to explore issues raised by each rule, including what counts as students' own work, how much AI assistance is acceptable, how AI use should be documented, and how suspected violations are handled. Because participants might be reluctant to discuss their GenAI practices, we emphasized that there were no right or wrong answers, that responses would be de-identified, and that nothing would be shared with their instructors or institutions. Interviews were conducted individually over Zoom and lasted approximately 45--60 minutes. We obtained verbal consent before each interview and recorded it with permission for transcription. Participants received \$20 checks as gifts for their time. We conducted thematic analysis~\cite{BraunClarke2006}, first deductively organizing transcripts by the rules being discussed, then inductively identifying patterns in how participants understood and experienced those rules. Two researchers coded the transcripts independently and met regularly to compare codes, resolve disagreements through discussion, and group codes into themes. After the initial coding stage, we used affinity diagramming to further add, remove, or merge themes. The full research team also participated at this stage in refining the themes iteratively.

\subsection{Findings}
\label{sec:findings}

Teachers and students did not discuss the same rules. \textbf{AI Disclosure} (R5) came up most often among teachers (13 out of 20), followed by \textbf{Instructor Approval} (R1) and \textbf{Content Prohibition} (R3), each
discussed by 10 out of 20 teachers, and \textbf{Misrepresented Authorship} (R4)  and \textbf{Academic Penalties} (R9), each discussed by 9 out of 20. Students also discussed \textbf{AI Disclosure} (R5) most often (12 out of 20), followed by \textbf{Specified AI Uses} (R8), discussed by 10 out of 20 students, and
\textbf{Content Prohibition} (R3), discussed by 9 out of 20. \textbf{Data Confidentiality} (R7) drew the least attention from both groups, discussed by only 2 out of 20 teachers and 5 out of 20 students. These differences suggest that despite a shared emphasis on \textbf{AI Disclosure} (R5), teachers and students prioritize different aspects of institutional GenAI Policy. Below, we describe how the policies shaped teachers' and students' perceptions and practices around GenAI.

\subsubsection{Compliance depends on perceived consequences and enforcement of policies}
\label{sec:find-negotiating}
Students described how their decisions to comply with GenAI policies were influenced by the perceived consequences of disclosure and the likelihood of enforcement. When acknowledging GenAI use was perceived as potentially affecting their grades, some students preferred to conceal their use. Discussing \textbf{Misrepresented Authorship} (R4), S17 stated, \textit{``I pretend not to use it, I pretend that I generate everything by myself.''} Others continued using GenAI despite explicit prohibitions, particularly when they believed their use would be difficult to detect. For example, S02 recalled a course that banned GenAI under \textbf{Content Prohibition} (R3) but had no apparent enforcement mechanism, stating, \textit{``We don't want you to use it. Yeah, there's no consequences, I guess.''} Similarly, asked whether students would follow a complete prohibition, S07 answered \textit{``no,''} having seen \textit{``peers using it and getting away with it, and even getting better grades.''} Conversely, S10 reported assuming that instructors might use AI-detection tools under \textbf{Academic Penalties} (R9) despite being uncertain whether such tools were actually employed.

\subsubsection{Mismatch between Personal Judgment and Policy Requirement}
\label{sec:find-mismatch}
Noncompliance did not necessarily reflect students' rejection of academic integrity expectations. S05 acknowledged using GenAI in a course that prohibited it but emphasized that they avoided directly copying generated content and ensured the submitted work remained their own:
 
\begin{quote}
\textit{``I have taken one, yes. And, like, to be honest, I still did use it. [...] I just tried to be, like, intentional, like, I would never, like, copy and paste anything, like, it was still all my own.''}
\end{quote}

Several rules leave students to interpret the boundaries of AI Use. Students perceived R1 (\textbf{Instructor Approval}), R2 (\textbf{Sole Ownership}), R4 (\textbf{Misrepresented Authorship}), and R9 (\textbf{Academic Penalties}) as vague, particularly because syllabus statements often failed to specify the boundaries of acceptable AI use. S04 suggested that instructors may intentionally leave these boundaries undefined, recognizing that students are likely to use AI regardless. In the absence of clear guidance, students established their own interpretations of acceptable use, often distinguishing between directly copying AI-generated content and using AI as a tool while retaining ownership of their work. As S03 explained, \textit{“The difference is taking those ideas and turning them into real thought, the what, the why, and the how, instead of just AI, which is giving you the what.”} This suggests that students sometimes relied on personal judgments about acceptable assistance when their practices conflicted with formal restrictions or when the policies were vague.

\subsubsection{Social stigma is a barrier to following policies}
\label{sec:find-stigma}

Students’ willingness to follow AI policies was shaped not only by the rules themselves but also by the stigma surrounding AI use. Rules that framed AI use as misconduct, such as \textbf{Misrepresented Authorship} (R4) and \textbf{Academic Penalties} (R9), added to students’ concerns about being judged or facing consequences (S03, S06, S10, S17). S03 explained how repeated discouragement left them unsure whether using AI would get them in trouble:
\begin{quote}
    \textit{“It feels like they kind of tell us always, don’t use AI, so it is kind of, is this gonna end up getting me in trouble? Because everyone preaches constantly not to use AI. I think there’s a stigmatism against it.”}
\end{quote}

In contrast, where AI use was openly accepted, students described attributing it as a routine step. S03 recalled an instructor who permitted AI use and asked only for attribution: \textit{“He encouraged the use of AI. He said, just make sure you’re [...] making the citations correctly.”} Paired with clear attribution expectations under \textbf{AI Disclosure} (R5), this acceptance made acknowledging AI use feel like ordinary practice rather than an admission. Teachers who took this stance also treated unreported use lightly, T08 did not consider it \textit{“a big deal”} in their courses. Thus, stigma discouraged students from acknowledging AI use, while acceptance paired with clear expectations for attribution and accountability made following the rules part of ordinary coursework.

\subsubsection{Explicit encouragement of accountability raises students' sense of responsibility}
\label{sec:find-accountability}

\textbf{Output Accountability} (R6), which holds students responsible for errors,
bias, or unethical AI output, drew comment from teachers across disciplines
(T03, T06, T09, T12, T13, T17, T19, T20), though they differed on who carries
that responsibility. Some placed it solely on the student. T13 already used
the rule in their classes and graded whatever a student submitted, regardless
of whether an error came from AI:
 
\begin{quote}
\textit{``You're responsible for what you submit. If you use AI to help you
write it, and it gets something wrong [...] you're going to get a grade based
on what you submit.''}
\end{quote}
 
T17 framed the same expectation as being able to defend the information in
one's work. A few teachers saw the responsibility as shared. T06 agreed that
students are accountable, but argued that while the tools are new, teachers
must show students what misuse looks like. For some teachers, the rule was less of a warning than a lesson in how to use
AI well. As T09 stated, 
 
\begin{quote}
\textit{``Some students, they actually came to my office hour asking me to
clarify some questions because they are kind of misled by some answers
provided by the AI. So I think rule number six kind of encourages the students
to not only use AI, but also use AI wisely.''}
\end{quote}

Students largely accepted this responsibility, and several described checking
AI output as their own habit (S01, S02, S06, S07, S12). S12 tied it to
submission: \textit{``it falls back on you, because you're the one submitting it
at the end of the day.''}

\subsubsection{Fear of unfair treatment among students}
\label{sec:find-unfair}
Rules like \textbf{Instructor Approval} (R1), \textbf{Misrepresented Authorship} (R4), and \textbf{Academic Penalties} (R9) led some students to expect that following the rules would
result in unfair outcomes. For example, \textbf{Instructor Approval} (R1) carried a risk for students, because asking for permission can draw attention to oneself and possibly to classmates. S04 chose not
to ask, explaining:

\begin{quote}
 \textit{``One, I don't want to be a tattletale, but two, I don't want to find out
that I'm not supposed to be using it, and everyone else is using it.''}
\end{quote}

Under \textbf{Misrepresented Authorship} (R4), S17
expected instructors to view students who disclosed their AI use as \textit{``lazy''} and give them a lower score. S05 expected that an instructor \textit{``would be more critical in grading.''}

\subsubsection{Enforcement requires trust and careful manual effort}
\label{sec:find-verifying}
Teachers mentioned enforcement to be a recurrent problem.
 T15 characterized enforcement under \textbf{Academic Penalties} (R9) as a matter of trust, noting that institutional restrictions on AI-detection tools and the lack of reliable verification methods made it difficult to determine whether students followed policy requirements:
 
\begin{quote}
\textit{``I know that at our university, we're not supposed to be checking, like, using AI detectors or whatever [...] there's not great ways to verify, right, if my students are following my syllabus or not, right? And so, it is just sort of a trust situation.''}
\end{quote}
 
Similarly, T13 questioned the accuracy of AI-detection tools, stating, \textit{``I'd have to be very confident, and I don't think I have that confidence.''} These concerns limited instructors' ability to substantiate suspected violations and complicated decisions about academic misconduct. Without tools they trusted, some instructors relied on their own judgment instead. T11 explained, \textit{``I don't use any of those formal tools. I just rely on my perceptions,''} adding that they gave students the benefit of the doubt when a case was ambiguous. Similarly while discussing  \textbf{Sole Ownership} (R2), T08 found it \textit{``really difficult [...] to tell whether they use AI or not''} and instead judged answers on their merits, while acknowledging that their own sense of what AI-generated text looks like might be biased. Under \textbf{Content Prohibition} (R3), T04 described encountering different responses when confronting students about suspected AI-generated submissions: some admitted using GenAI, others denied it, and some accepted a failing grade without responding.
 
Instructors also described the time and effort that enforcing these rules would require, particularly in large classes. T03 viewed \textbf{Instructor Approval} (R1) as reasonable in principle but impractical at scale, explaining that it \textit{``only puts one more layer of communication between the faculties and the students, specifically if you have a class with 200 students.''} For \textbf{Content Prohibition} (R3), T13 had observed colleagues who adopted a complete ban spend considerable time trying to identify AI use, meeting with students, and reporting cases to the office of academic integrity. For these instructors, the burden of enforcement made strict rules difficult to sustain in practice.

\subsubsection{Disclosure is the easiest and least stressful attribution mechanism}
There is a reason why \textbf{AI Disclosure} (R5) was the most popular rule among both students and teachers. Students (S03, S09, S10, S13) mentioned that writing a disclosure is the easiest way to attribute AI use compared to situations under \textbf{Misrepresented Authorship} (R4), where they had to manually mark AI-generated texts or submit a complete chat history. Teachers valued proper disclosures and said that most of the time, they are enough to check whether a student is honest or not.

A caveat here is that ``easy'' might not always result into ``appropriate'' attribution. Students and teachers rely on disclosure because it is comparatively easier to create and evaluate. However, several teachers(T03, T04, T08, T12) mentioned that disclosure can reduce many legitimate and illegitimate uses of GenAI to a simple sentence, which is uninformative in some cases. In T08's words, \textit{``it's just not enough to say I use the tool.''} Instructors also
needed to understand what it was used for and how it contributed to the work.

\subsubsection{Signs of critical thinking outweigh any rule}
\label{sec:find-critical}
A recurrent theme in the interviews was that policies should align with learning outcomes.  Several teachers did not object to students using AI in coursework, provided
the work showed that the student had done the thinking (T05, T06, T20). For
these teachers, critical thinking, not compliance with a particular rule, was
the standard by which AI use should be judged. T06 mentioned:

\begin{quote}
\textit{``As long as there is signature of you thinking critically. I'm fine.
[...] now the question is, how do you identify whether the student has
thought critically or not?''}
\end{quote}

Teachers' acceptance of rules were shaped by course's learning outcomes.
Where a permitted use of GenAI let students skip the thinking a course was meant to
build, teachers withdrew the permission. T12 explained that, in a reading and
writing course, using AI to summarize assigned texts or to write on a
student's behalf falls outside \textbf{Specified AI Uses} (R8) because it ``shortcut[s]''
the very skills being taught. T15 made the dependence on outcomes explicit:
\textbf{Content Prohibition} (R3) suited courses where AI worked against the learning
goals, while courses preparing students for the workplace should teach
effective use instead. Students shared a similar view, where they want to optimize learning skills, rather than following rules (S02, S03, S05, S09, S10, S12, S17).

\subsection{{Policy-Impact Matrix}}

\begin{figure}
    \centering
    \includegraphics[width=\linewidth]{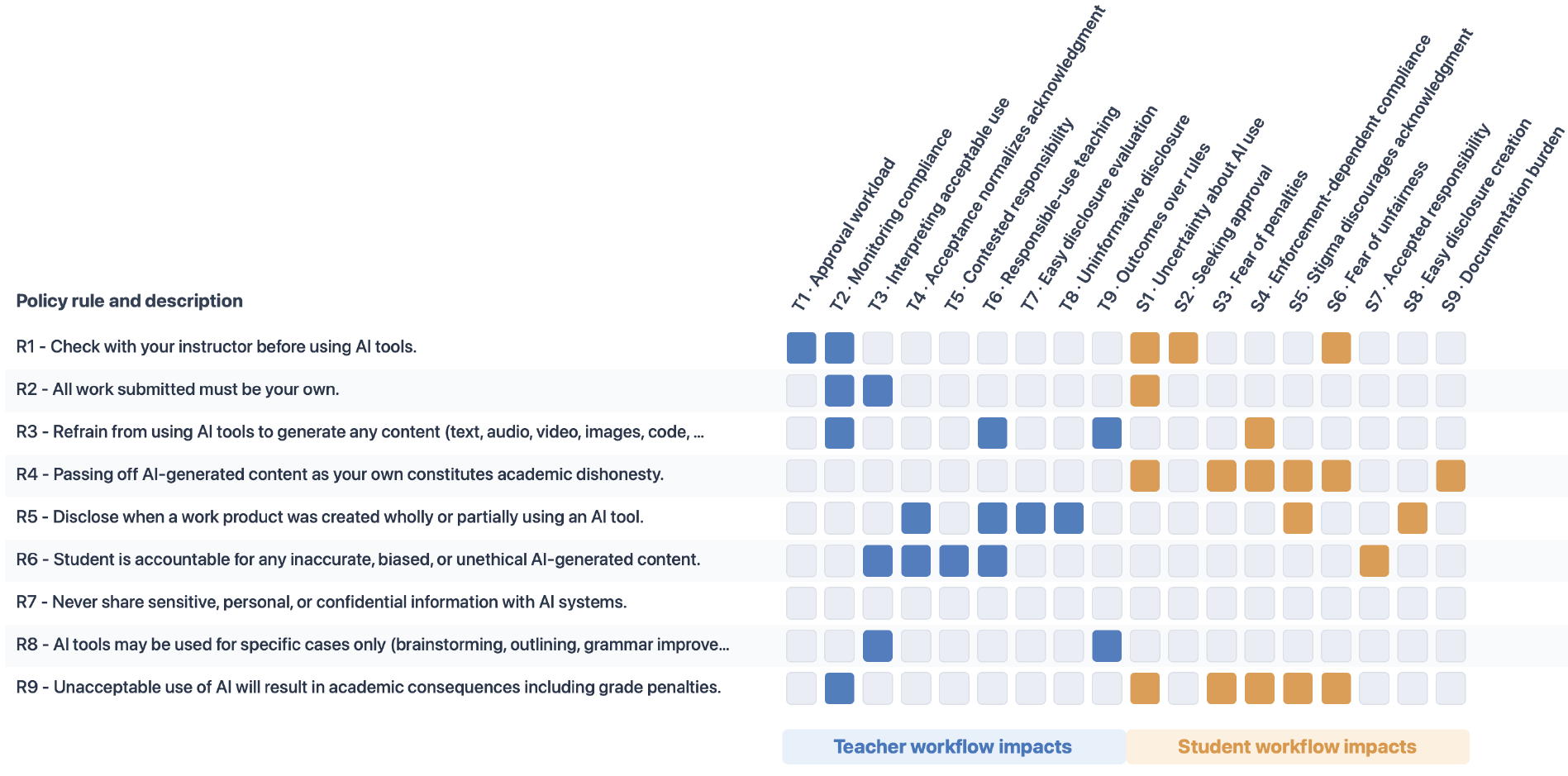}
    \caption{\textbf{Policy-Impact Matrix}. Vertical axis represents the nine rules, while the horizontal axis represents impact of policies on teacher and student workflows. Filled cells denote recorded relationships; all relationships have equal weight.}
    \label{fig:matrix}
\end{figure}

We summarize our findings in a policy-impact matrix (Figure~\ref{fig:matrix}). The matrix summarizes how the policy rules can impact the workflows and perspectives of both teachers and students. We see the matrix as a resource for teachers and educators to understand the practical implications of the nine rules. Students can use the matrix to understand the requirements of the tools, but the primary users are teachers since they decide how the rules are adopted and applied in classrooms. Several interesting insights arise from the matrix. For example, R7 (Data Confidentiality) has no impacts since participants in the study did not find it relevant for the classes. R1, which was the most popular rule in our policy corpus, added approval workload (T1) as teachers needed to oversee each request from students. Fear of unfair outcomes (S6) was related to rules with negative connotations (R1, R4, R9). Additionally, R4 and R9, both of which mention academic dishonesty and penalties, resulted into more workflow impacts for students compared to other rules, indicating that these rules are stressful for students to follow. In contrast, the constructive impacts for students, accepted responsibility (S7) and easy disclosure creation (S8), appear only under R5 and R6, the rules that ask students to account for their AI use rather than avoid it An interactive web-based version of the matrix can be found  \href{https://osf.io/23z4n/overview?view_only=6f8ed1d0e2864953a3312dfe65d14332}{here}. Upon acceptance, we plan to release the interactive matrix as a public and interactive tool for navigating AI-based policies for teachers and students.

\section{Discussion}\label{sec:discussion}

We see our work as a call to investigate GenAI policies and educational outcomes more closely. The impacts that policies can have on students and teachers are nuanced but significant. Even a simple wording difference in the policies can significantly shape how teachers and students act upon them. 
However, educational institutes have not fully realized these impacts, which could significantly change how the policies are developed, what they require, and the support that institutes provide to teachers and students. Our work is a starting point in that direction. In the following, we outline two possible research directions that can further bridge the gap between GenAI policies and educational outcomes.

\subsection{Developing, coordinating, and communicating policies via structured forms}\label{sec:discussion:developing}
Our research shows that current GenAI policies leave instructors and students with substantial interpretive work --- \ie instructors need to translate broad institutional and course-specific rules into expectations for specific courses or assignments, while students need to determine how their GenAI use fits these expectations.
Research on documentation and disclosure suggests that structured forms could support this work.
Kelley \etal \cite{kelley-chi2010} found that standardized privacy notices improved the ability for participants to locate relevant policy information.
Work on privacy disclosures argues that this structured reporting can establish clear benchmarks for assessing compliance \cite{JOLT-2023}, while filling in these forms can help organizations communicate information more effectively across units \cite{JOLT-2025}.
A similar idea is also proposed by Gebru \etal \cite{gebru-cacm2021} to support communication between the producers and consumers of datasets.
These perspectives support the idea that structured forms can broadly help communication regarding the expectations of GenAI across all levels.
At the institutional level, these shared forms could require policy authors to specify their requirements succinctly, whom they apply to, and which decisions are left to other stakeholders.
Reviewing these responses together could help institutions identify overlaps, resolve inconsistencies in expectations, and clarify roles.
The single form resulting from the aggregated inputs can help faculty can more easily distinguish the boundaries of their discretion (\ie what are institution-wide expectations and what are the boundaries within which their course-level expectations may reside).
At the course-level, instructors could append their expectations in similarly succinct and structured forms to specify which specific activities are permitted, what must be disclosed and how, how these choices relate to the course objectives, and how expectations will be enforced; and students could attest their adherence to the institution-wide and course-specific expectations in their submitted work.

In general, such structured forms would simultaneously: 
(1) connect the submitted work from students with expectations communicated by faculty and institutions;
(2) make expectations easier to locate and compare across courses through the use of common fields and explanation types (further, our rule cards could inform decisions about wording in these fields or their explanations by helping designers anticipate the interpretive and practical challenges associated with different expectations); and 
(3) support discussion at each level, allowing unresolved questions or specific use cases to reach policy authors and inform revisions.
Of course, their effectiveness would depend on whether they improve shared understanding without imposing excessive administrative burdens on stakeholders.

\subsection{Tools for more effective enforcement of policies}\label{sec:discussion:enforcing}

A recurrent theme in the interviews is that teachers struggle to enforce some policy rules because they do not have an effective way to measure violations. Similarly, students are often uncertain about the evidence they need to submit to show compliance. 

One direction to address this gap is to develop tools that connect GenAI use policies with records of students' interactions with GenAI.
In such tools, instructors could select and customize applicable rules when creating assignments (\eg using the structured forms described in \Cref{sec:discussion:developing}).
Students could then work in an environment with integrated GenAI assistance that records all relevant interactions and automatically connects them with the resulting submission~\cite{hoque2024chi}.
Such tools would allow context for assessing the use against the stated expectations, while leaving judgment to the instructor.
Where potential conflicts between GenAI use and course expectations are identified, the tool could prompt students to seek clarification from the instructor.
Future work might investigate the feasibility of such tools and whether they support more consistent assessment, encourage policy dialogue, effectively manage documentation, and mitigate burdens on instructors while supporting students rather than making them feel anxious or surveilled.

\section{Conclusion}
\label{sec:conclusion}

We analyzed policies from 101 U.S.\ universities to derive nine
GenAI provenance rules, then interviewed 20 instructors and 20 students about
what those rules ask of them. The same wording produces incompatible
instructions from instructors in the same discipline, and produces disclosure
or suppression in students depending on whether it is framed as an invitation
or a sanction. Policies are more specific about what is forbidden than about
what would demonstrate compliance. Our policy-impact matrix shows how different rules can have similar effects. It helps stakeholders understand how these rules may affect teachers' and students' workflows before adopting them. It also helps stakeholders revise related rules together rather than separately.

\section*{Generative AI Disclosure Statement}
Generative AI was part of our policy-analysis method (Section 3). AI agents driven by OpenAI’s GPT6 Astra were used for policy collection (Figure 4) and analysis (Figure 5), guided by author-written inclusion criteria, a nine-rule codebook, and 432 hand-coded examples. The authors wrote all prompts, and one author manually reviewed 320 sampled excerpts to validate the output (Section 3.3). Interviews were coded by two authors without generative AI. We used Claude Sonnet 5.5 (Anthropic) to restructure parts of our writing and for grammar correction. The authors reviewed all AI-edited text. All findings, interpretations, and conclusions are the authors’, who take full responsibility for the paper’s content.

\section*{Ethical Considerations Statement}
The policy analysis used publicly available institutional documents and did not involve human participants. The interview studies were reviewed by our
institution's Institutional Review Board. Participants gave informed consent before the interview, including consent to audio recording. They were told that they could decline any question or stop at any time. 
This study posed minimal risk to participants. The main concern was confidentiality, because students described how they used GenAI in their coursework and what they reported to their instructors, and teachers described cases in which they suspected undisclosed use. We therefore did not collect course identifiers, removed identifying details from transcripts, and refer to participants only by pseudonymous identifiers (T01--T20 and S01--S20). We report institution and discipline only at a level of detail that does not allow participants to be re-identified. Audio recordings were stored on secured  institutional storage and is only accessible to authorized researchers. No identifiable participant data were entered into third-party Generative AI tools. The materials we shared publicly also do not include any identifiable information of the participants. The authors declare no conflicts of interest related to this research.

\bibliographystyle{ACM-Reference-Format}
\bibliography{main,chi_references}


\begin{thebibliography}{44}


\ifx \showCODEN    \undefined \def \showCODEN     #1{\unskip}     \fi
\ifx \showISBNx    \undefined \def \showISBNx     #1{\unskip}     \fi
\ifx \showISBNxiii \undefined \def \showISBNxiii  #1{\unskip}     \fi
\ifx \showISSN     \undefined \def \showISSN      #1{\unskip}     \fi
\ifx \showLCCN     \undefined \def \showLCCN      #1{\unskip}     \fi
\ifx \shownote     \undefined \def \shownote      #1{#1}          \fi
\ifx \showarticletitle \undefined \def \showarticletitle #1{#1}   \fi
\ifx \showURL      \undefined \def \showURL       {\relax}        \fi
\providecommand\bibfield[2]{#2}
\providecommand\bibinfo[2]{#2}
\providecommand\natexlab[1]{#1}
\providecommand\showeprint[2][]{arXiv:#2}

\bibitem[Adnin et~al\mbox{.}(2025)]%
        {adnin_examining_2025}
\bibfield{author}{\bibinfo{person}{Rudaiba Adnin}, \bibinfo{person}{Atharva Pandkar}, \bibinfo{person}{Bingsheng Yao}, \bibinfo{person}{Dakuo Wang}, {and} \bibinfo{person}{Maitraye Das}.} \bibinfo{year}{2025}\natexlab{}.
\newblock \showarticletitle{Examining {Student} and {Teacher} {Perspectives} on {Undisclosed} {Use} of {Generative} {AI} in {Academic} {Work}}. In \bibinfo{booktitle}{\emph{Proceedings of the 2025 {CHI} {Conference} on {Human} {Factors} in {Computing} {Systems}}} \emph{(\bibinfo{series}{{CHI} '25})}. \bibinfo{publisher}{Association for Computing Machinery}, \bibinfo{address}{New York, NY, USA}, \bibinfo{pages}{1--17}.
\newblock
\showISBNx{979-8-4007-1394-1}
\href{https://doi.org/10.1145/3706598.3713393}{doi:\nolinkurl{10.1145/3706598.3713393}}


\bibitem[An et~al\mbox{.}(2025)]%
        {an2025investigating}
\bibfield{author}{\bibinfo{person}{Yunjo An}, \bibinfo{person}{Ji~Hyun Yu}, {and} \bibinfo{person}{Shadarra James}.} \bibinfo{year}{2025}\natexlab{}.
\newblock \showarticletitle{Investigating the higher education institutions’ guidelines and policies regarding the use of generative AI in teaching, learning, research, and administration}.
\newblock \bibinfo{journal}{\emph{International Journal of Educational Technology in Higher Education}} \bibinfo{volume}{22}, \bibinfo{number}{1} (\bibinfo{year}{2025}), \bibinfo{pages}{10}.
\newblock


\bibitem[Braun and Clarke(2006)]%
        {BraunClarke2006}
\bibfield{author}{\bibinfo{person}{Virginia Braun} {and} \bibinfo{person}{Victoria Clarke}.} \bibinfo{year}{2006}\natexlab{}.
\newblock \showarticletitle{Using Thematic Analysis in Psychology}.
\newblock \bibinfo{journal}{\emph{Qualitative Research in Psychology}} \bibinfo{volume}{3}, \bibinfo{number}{2} (\bibinfo{year}{2006}), \bibinfo{pages}{77--101}.
\newblock
\href{https://doi.org/10.1191/1478088706qp063oa}{doi:\nolinkurl{10.1191/1478088706qp063oa}}


\bibitem[Chan(2023)]%
        {chan2023comprehensive}
\bibfield{author}{\bibinfo{person}{Cecilia Ka~Yuk Chan}.} \bibinfo{year}{2023}\natexlab{}.
\newblock \showarticletitle{A comprehensive AI policy education framework for university teaching and learning}.
\newblock \bibinfo{journal}{\emph{International journal of educational technology in higher education}} \bibinfo{volume}{20}, \bibinfo{number}{1} (\bibinfo{year}{2023}), \bibinfo{pages}{38}.
\newblock


\bibitem[Corbin et~al\mbox{.}(2025)]%
        {corbin2025talk}
\bibfield{author}{\bibinfo{person}{Thomas Corbin}, \bibinfo{person}{Phillip Dawson}, {and} \bibinfo{person}{Danny Liu}.} \bibinfo{year}{2025}\natexlab{}.
\newblock \showarticletitle{Talk Is Cheap: Why Structural Assessment Changes Are Needed for a Time of {GenAI}}.
\newblock \bibinfo{journal}{\emph{Assessment \& Evaluation in Higher Education}} \bibinfo{volume}{50}, \bibinfo{number}{7} (\bibinfo{year}{2025}), \bibinfo{pages}{1087--1097}.
\newblock
\href{https://doi.org/10.1080/02602938.2025.2503964}{doi:\nolinkurl{10.1080/02602938.2025.2503964}}


\bibitem[Dabis and Cs{\'a}ki(2024)]%
        {dabis2024ai}
\bibfield{author}{\bibinfo{person}{Attila Dabis} {and} \bibinfo{person}{Csaba Cs{\'a}ki}.} \bibinfo{year}{2024}\natexlab{}.
\newblock \showarticletitle{AI and ethics: Investigating the first policy responses of higher education institutions to the challenge of generative AI}.
\newblock \bibinfo{journal}{\emph{Humanities and Social Sciences Communications}} \bibinfo{volume}{11}, \bibinfo{number}{1} (\bibinfo{year}{2024}), \bibinfo{pages}{1006}.
\newblock


\bibitem[Dai(2025)]%
        {dai_why_2025}
\bibfield{author}{\bibinfo{person}{Yun Dai}.} \bibinfo{year}{2025}\natexlab{}.
\newblock \showarticletitle{Why students use or not use generative {AI}: {Student} conceptions, concerns, and implications for engineering education}.
\newblock \bibinfo{journal}{\emph{Digital Engineering}}  \bibinfo{volume}{4} (\bibinfo{date}{March} \bibinfo{year}{2025}), \bibinfo{pages}{100019}.
\newblock
\showISSN{2950-550X}
\href{https://doi.org/10.1016/j.dte.2024.100019}{doi:\nolinkurl{10.1016/j.dte.2024.100019}}


\bibitem[Diamantis et~al\mbox{.}(2023)]%
        {JOLT-2023}
\bibfield{author}{\bibinfo{person}{Mihailis Diamantis}, \bibinfo{person}{Maaz~Bin Musa}, \bibinfo{person}{Lucas Ausberger}, {and} \bibinfo{person}{Rishab Nithyanand}.} \bibinfo{year}{2023}\natexlab{}.
\newblock \showarticletitle{{Forms of Disclosure: The Path to Automated Data Privacy Audits}}. In \bibinfo{booktitle}{\emph{{Harvard Journal of Law and Technology}}}.
\newblock
\urldef\tempurl%
\url{https://heinonline.org/HOL/P?h=hein.journals/hjlt37&i=531}
\showURL{%
\tempurl}


\bibitem[Diamantis et~al\mbox{.}(2026)]%
        {JOLT-2025}
\bibfield{author}{\bibinfo{person}{Mihailis~E Diamantis}, \bibinfo{person}{Chen Sun}, {and} \bibinfo{person}{Rishab Nithyanand}.} \bibinfo{year}{2026}\natexlab{}.
\newblock \showarticletitle{Information About Data}.
\newblock \bibinfo{journal}{\emph{Yale JL \& Tech.}}  \bibinfo{volume}{28} (\bibinfo{year}{2026}), \bibinfo{pages}{238}.
\newblock


\bibitem[Gebru et~al\mbox{.}(2021)]%
        {gebru-cacm2021}
\bibfield{author}{\bibinfo{person}{Timnit Gebru}, \bibinfo{person}{Jamie Morgenstern}, \bibinfo{person}{Briana Vecchione}, \bibinfo{person}{Jennifer~Wortman Vaughan}, \bibinfo{person}{Hanna Wallach}, \bibinfo{person}{Hal~Daum{\'e} III}, {and} \bibinfo{person}{Kate Crawford}.} \bibinfo{year}{2021}\natexlab{}.
\newblock \showarticletitle{Datasheets for datasets}.
\newblock \bibinfo{journal}{\emph{Commun. ACM}} \bibinfo{volume}{64}, \bibinfo{number}{12} (\bibinfo{date}{Nov.} \bibinfo{year}{2021}), \bibinfo{pages}{86–92}.
\newblock
\showISSN{0001-0782}
\href{https://doi.org/10.1145/3458723}{doi:\nolinkurl{10.1145/3458723}}


\bibitem[Gonsalves(2025)]%
        {gonsalves2025addressing}
\bibfield{author}{\bibinfo{person}{Chahna Gonsalves}.} \bibinfo{year}{2025}\natexlab{}.
\newblock \showarticletitle{Addressing student non-compliance in AI use declarations: Implications for academic integrity and assessment in higher education}.
\newblock \bibinfo{journal}{\emph{Assessment \& Evaluation in Higher Education}} \bibinfo{volume}{50}, \bibinfo{number}{4} (\bibinfo{year}{2025}), \bibinfo{pages}{592--606}.
\newblock


\bibitem[Gorichanaz(2023)]%
        {gorichanaz2023accused}
\bibfield{author}{\bibinfo{person}{Tim Gorichanaz}.} \bibinfo{year}{2023}\natexlab{}.
\newblock \showarticletitle{Accused: How Students Respond to Allegations of Using {ChatGPT} on Assessments}.
\newblock \bibinfo{journal}{\emph{Learning: Research and Practice}} \bibinfo{volume}{9}, \bibinfo{number}{2} (\bibinfo{year}{2023}), \bibinfo{pages}{183--196}.
\newblock
\href{https://doi.org/10.1080/23735082.2023.2254787}{doi:\nolinkurl{10.1080/23735082.2023.2254787}}


\bibitem[Hoque et~al\mbox{.}(2024)]%
        {hoque2024chi}
\bibfield{author}{\bibinfo{person}{Naimul Hoque}, \bibinfo{person}{Tasfia Mashiat}, \bibinfo{person}{Bhavya Ghai}, \bibinfo{person}{Cecilia~D. Shelton}, \bibinfo{person}{Fanny Chevalier}, \bibinfo{person}{Kari Kraus}, {and} \bibinfo{person}{Niklas Elmqvist}.} \bibinfo{year}{2024}\natexlab{}.
\newblock \showarticletitle{The HaLLMark Effect: Supporting Provenance and Transparent Use of Large Language Models in Writing with Interactive Visualization}. In \bibinfo{booktitle}{\emph{Proceedings of the 2024 CHI Conference on Human Factors in Computing Systems}}. \bibinfo{publisher}{ACM}, \bibinfo{address}{New York, NY, USA}, Article \bibinfo{articleno}{1045}, \bibinfo{numpages}{15}~pages.
\newblock
\href{https://doi.org/10.1145/3613904.3641895}{doi:\nolinkurl{10.1145/3613904.3641895}}


\bibitem[Hosseini et~al\mbox{.}(2025)]%
        {Hosseini2025}
\bibfield{author}{\bibinfo{person}{Mohammad Hosseini}, \bibinfo{person}{Bert Gordijn}, \bibinfo{person}{Gregory~E. Kaebnick}, {and} \bibinfo{person}{Kristi Holmes}.} \bibinfo{year}{2025}\natexlab{}.
\newblock \showarticletitle{Disclosing generative AI use for writing assistance should be voluntary}.
\newblock \bibinfo{journal}{\emph{Research Ethics}} \bibinfo{volume}{21}, \bibinfo{number}{4} (\bibinfo{year}{2025}), \bibinfo{pages}{728--735}.
\newblock
\href{https://doi.org/10.1177/17470161251345499}{doi:\nolinkurl{10.1177/17470161251345499}}


\bibitem[Jiang and Nakatani(2025)]%
        {jiang_exploring_2025}
\bibfield{author}{\bibinfo{person}{Yabing Jiang} {and} \bibinfo{person}{Kazuo Nakatani}.} \bibinfo{year}{2025}\natexlab{}.
\newblock \showarticletitle{Exploring {Implementations} of {GenAI} in {Teaching} {IS} {Subjects} and {Student} {Perceptions}}.
\newblock \bibinfo{journal}{\emph{Journal of Information Systems Education}} \bibinfo{volume}{36}, \bibinfo{number}{2} (\bibinfo{date}{June} \bibinfo{year}{2025}), \bibinfo{pages}{180--194}.
\newblock
\showISSN{2574-3872}
\href{https://doi.org/10.62273/WFHO1011}{doi:\nolinkurl{10.62273/WFHO1011}}


\bibitem[Jin et~al\mbox{.}(2025)]%
        {jin2025generative}
\bibfield{author}{\bibinfo{person}{Yueqiao Jin}, \bibinfo{person}{Lixiang Yan}, \bibinfo{person}{Vanessa Echeverria}, \bibinfo{person}{Dragan Ga{\v{s}}evi{\'c}}, {and} \bibinfo{person}{Roberto Martinez-Maldonado}.} \bibinfo{year}{2025}\natexlab{}.
\newblock \showarticletitle{Generative {AI} in Higher Education: A Global Perspective of Institutional Adoption Policies and Guidelines}.
\newblock \bibinfo{journal}{\emph{Computers and Education: Artificial Intelligence}}  \bibinfo{volume}{8} (\bibinfo{year}{2025}), \bibinfo{pages}{100348}.
\newblock
\href{https://doi.org/10.1016/j.caeai.2024.100348}{doi:\nolinkurl{10.1016/j.caeai.2024.100348}}


\bibitem[Kaufmann et~al\mbox{.}(2026)]%
        {kaufmann2026trend}
\bibfield{author}{\bibinfo{person}{Renee Kaufmann}, \bibinfo{person}{Jessalyn~I Vallade}, \bibinfo{person}{Brandi~N Frisby}, {and} \bibinfo{person}{Joe~C Martin}.} \bibinfo{year}{2026}\natexlab{}.
\newblock \showarticletitle{A trend examination of course policy and message framing for student GenAI}.
\newblock \bibinfo{journal}{\emph{Interactive Learning Environments}} \bibinfo{volume}{34}, \bibinfo{number}{6} (\bibinfo{year}{2026}), \bibinfo{pages}{3995--4011}.
\newblock


\bibitem[Kelley et~al\mbox{.}(2010)]%
        {kelley-chi2010}
\bibfield{author}{\bibinfo{person}{Patrick~Gage Kelley}, \bibinfo{person}{Lucian Cesca}, \bibinfo{person}{Joanna Bresee}, {and} \bibinfo{person}{Lorrie~Faith Cranor}.} \bibinfo{year}{2010}\natexlab{}.
\newblock \showarticletitle{Standardizing privacy notices: an online study of the nutrition label approach}. In \bibinfo{booktitle}{\emph{Proceedings of the SIGCHI Conference on Human Factors in Computing Systems}} (Atlanta, Georgia, USA) \emph{(\bibinfo{series}{CHI '10})}. \bibinfo{publisher}{Association for Computing Machinery}, \bibinfo{address}{New York, NY, USA}, \bibinfo{pages}{1573–1582}.
\newblock
\showISBNx{9781605589299}
\href{https://doi.org/10.1145/1753326.1753561}{doi:\nolinkurl{10.1145/1753326.1753561}}


\bibitem[Kim et~al\mbox{.}(2026)]%
        {kim2026attribution}
\bibfield{author}{\bibinfo{person}{W. Kim}, \bibinfo{person}{C. Lee}, {and} \bibinfo{person}{H. Kim}.} \bibinfo{year}{2026}\natexlab{}.
\newblock \showarticletitle{An Attribution Framework for Transparent Use of Generative {AI} in University Coursework}.
\newblock \bibinfo{journal}{\emph{The Journal of Korean Association of Computer Education}} \bibinfo{volume}{29}, \bibinfo{number}{3} (\bibinfo{year}{2026}), \bibinfo{pages}{20--28}.
\newblock
\href{https://doi.org/10.32431/kace.2026.29.3.003}{doi:\nolinkurl{10.32431/kace.2026.29.3.003}}


\bibitem[LaFrance(2026)]%
        {lafrance2026governing}
\bibfield{author}{\bibinfo{person}{Jason LaFrance}.} \bibinfo{year}{2026}\natexlab{}.
\newblock \showarticletitle{Governing Generative Artificial Intelligence: Institutional Policies and Guidelines at America’s Flagship Universities}.
\newblock \bibinfo{journal}{\emph{Educational Policy}} (\bibinfo{year}{2026}), \bibinfo{pages}{08959048261450995}.
\newblock


\bibitem[Lau et~al\mbox{.}(2026)]%
        {lau2026barriers}
\bibfield{author}{\bibinfo{person}{Sam Lau}, \bibinfo{person}{Kianoosh Boroojeni}, \bibinfo{person}{Harry Keeling}, {and} \bibinfo{person}{Jenn Marroquin}.} \bibinfo{year}{2026}\natexlab{}.
\newblock \showarticletitle{Barriers that Programming Instructors Face While Performing Emergency Pedagogical Design to Shape Student-{AI} Interactions with Generative {AI} Tools}. In \bibinfo{booktitle}{\emph{Proceedings of the 2026 CHI Conference on Human Factors in Computing Systems}} \emph{(\bibinfo{series}{CHI '26})}. \bibinfo{publisher}{Association for Computing Machinery}, \bibinfo{address}{New York, NY, USA}, Article \bibinfo{articleno}{528}, \bibinfo{numpages}{15}~pages.
\newblock
\href{https://doi.org/10.1145/3772318.3790682}{doi:\nolinkurl{10.1145/3772318.3790682}}


\bibitem[Liang et~al\mbox{.}(2023)]%
        {liang_gpt_2023}
\bibfield{author}{\bibinfo{person}{Weixin Liang}, \bibinfo{person}{Mert Yuksekgonul}, \bibinfo{person}{Yining Mao}, \bibinfo{person}{Eric Wu}, {and} \bibinfo{person}{James Zou}.} \bibinfo{year}{2023}\natexlab{}.
\newblock \showarticletitle{{GPT} detectors are biased against non-native {English} writers}.
\newblock \bibinfo{journal}{\emph{Patterns}} \bibinfo{volume}{4}, \bibinfo{number}{7} (\bibinfo{date}{July} \bibinfo{year}{2023}), \bibinfo{pages}{100779}.
\newblock
\showISSN{26663899}
\href{https://doi.org/10.1016/j.patter.2023.100779}{doi:\nolinkurl{10.1016/j.patter.2023.100779}}


\bibitem[Lo(2023)]%
        {lo2023impact}
\bibfield{author}{\bibinfo{person}{Chung~Kwan Lo}.} \bibinfo{year}{2023}\natexlab{}.
\newblock \showarticletitle{What is the impact of ChatGPT on education? A rapid review of the literature}.
\newblock \bibinfo{journal}{\emph{Education sciences}} \bibinfo{volume}{13}, \bibinfo{number}{4} (\bibinfo{year}{2023}), \bibinfo{pages}{410}.
\newblock


\bibitem[Luo(2024)]%
        {luo2024critical}
\bibfield{author}{\bibinfo{person}{Jiahui Luo}.} \bibinfo{year}{2024}\natexlab{}.
\newblock \showarticletitle{A critical review of GenAI policies in higher education assessment: A call to reconsider the “originality” of students’ work}.
\newblock \bibinfo{journal}{\emph{Assessment \& Evaluation in Higher Education}} \bibinfo{volume}{49}, \bibinfo{number}{5} (\bibinfo{year}{2024}), \bibinfo{pages}{651--664}.
\newblock


\bibitem[Luo(2025)]%
        {luo2025does}
\bibfield{author}{\bibinfo{person}{Jiahui Luo}.} \bibinfo{year}{2025}\natexlab{}.
\newblock \showarticletitle{How does GenAI affect trust in teacher-student relationships? Insights from students’ assessment experiences}.
\newblock \bibinfo{journal}{\emph{Teaching in Higher Education}} \bibinfo{volume}{30}, \bibinfo{number}{4} (\bibinfo{year}{2025}), \bibinfo{pages}{991--1006}.
\newblock


\bibitem[McDonald et~al\mbox{.}(2025)]%
        {mcdonald2025generative}
\bibfield{author}{\bibinfo{person}{Nora McDonald}, \bibinfo{person}{Aditya Johri}, \bibinfo{person}{Areej Ali}, {and} \bibinfo{person}{Aayushi~Hingle Collier}.} \bibinfo{year}{2025}\natexlab{}.
\newblock \showarticletitle{Generative artificial intelligence in higher education: Evidence from an analysis of institutional policies and guidelines}.
\newblock \bibinfo{journal}{\emph{Computers in Human Behavior: Artificial Humans}}  \bibinfo{volume}{3} (\bibinfo{year}{2025}), \bibinfo{pages}{100121}.
\newblock


\bibitem[Moore and Lookadoo(2024)]%
        {moore2024communicating}
\bibfield{author}{\bibinfo{person}{Sarah Moore} {and} \bibinfo{person}{Kathryn Lookadoo}.} \bibinfo{year}{2024}\natexlab{}.
\newblock \showarticletitle{Communicating clear guidance: Advice for generative AI policy development in higher education}.
\newblock \bibinfo{journal}{\emph{Business and Professional Communication Quarterly}} \bibinfo{volume}{87}, \bibinfo{number}{4} (\bibinfo{year}{2024}), \bibinfo{pages}{610--629}.
\newblock


\bibitem[Moorhouse et~al\mbox{.}(2023)]%
        {MOORHOUSE2023100151}
\bibfield{author}{\bibinfo{person}{Benjamin~Luke Moorhouse}, \bibinfo{person}{Marie~Alina Yeo}, {and} \bibinfo{person}{Yuwei Wan}.} \bibinfo{year}{2023}\natexlab{}.
\newblock \showarticletitle{Generative AI tools and assessment: Guidelines of the world's top-ranking universities}.
\newblock \bibinfo{journal}{\emph{Computers and Education Open}}  \bibinfo{volume}{5} (\bibinfo{year}{2023}), \bibinfo{pages}{100151}.
\newblock
\showISSN{2666-5573}
\href{https://doi.org/10.1016/j.caeo.2023.100151}{doi:\nolinkurl{10.1016/j.caeo.2023.100151}}


\bibitem[P{\'e}rez-P{\'e}rez et~al\mbox{.}(2026)]%
        {perez-perez_transparency_2026}
\bibfield{author}{\bibinfo{person}{Itahisa P{\'e}rez-P{\'e}rez}, \bibinfo{person}{Miriam~Catalina Gonz{\'a}lez-Afonso}, \bibinfo{person}{Zeus Plasencia-Carballo}, {and} \bibinfo{person}{David P{\'e}rez-Jorge}.} \bibinfo{year}{2026}\natexlab{}.
\newblock \showarticletitle{Transparency {Mechanisms} for {Generative} {AI} {Use} in {Higher} {Education} {Assessment}: {A} {Systematic} {Scoping} {Review} (2022–2026)}.
\newblock \bibinfo{journal}{\emph{Computers}} \bibinfo{volume}{15}, \bibinfo{number}{2} (\bibinfo{date}{Feb.} \bibinfo{year}{2026}), \bibinfo{pages}{111}.
\newblock
\showISSN{2073-431X}
\href{https://doi.org/10.3390/computers15020111}{doi:\nolinkurl{10.3390/computers15020111}}


\bibitem[Petricini et~al\mbox{.}(2025)]%
        {petricini2025ai}
\bibfield{author}{\bibinfo{person}{Tiffany Petricini}, \bibinfo{person}{Sarah Zipf}, {and} \bibinfo{person}{Chuhao Wu}.} \bibinfo{year}{2025}\natexlab{}.
\newblock \showarticletitle{AI: Communicating academic honesty: Teacher messages and student perceptions about generative AI}.
\newblock \bibinfo{journal}{\emph{Frontiers in Communication}}  \bibinfo{volume}{10} (\bibinfo{year}{2025}), \bibinfo{pages}{1544430}.
\newblock


\bibitem[Qian(2026)]%
        {qian2026governing}
\bibfield{author}{\bibinfo{person}{Yufeng Qian}.} \bibinfo{year}{2026}\natexlab{}.
\newblock \showarticletitle{Governing generative AI in higher education: emerging policy approaches and support ecosystems at innovative US Universities}.
\newblock \bibinfo{journal}{\emph{International Journal for Educational Integrity}} \bibinfo{volume}{22}, \bibinfo{number}{1} (\bibinfo{year}{2026}), \bibinfo{pages}{25}.
\newblock


\bibitem[Qu et~al\mbox{.}(2025)]%
        {qu2025generative}
\bibfield{author}{\bibinfo{person}{Yao Qu}, \bibinfo{person}{Hui~En Loo}, {and} \bibinfo{person}{Jue Wang}.} \bibinfo{year}{2025}\natexlab{}.
\newblock \showarticletitle{Generative artificial intelligence in higher education: Emotional tensions and ethical declaration}.
\newblock \bibinfo{journal}{\emph{British Journal of Educational Technology}} (\bibinfo{year}{2025}).
\newblock


\bibitem[Sadasivan et~al\mbox{.}(2023)]%
        {sadasivan2023can}
\bibfield{author}{\bibinfo{person}{Vinu~Sankar Sadasivan}, \bibinfo{person}{Aounon Kumar}, \bibinfo{person}{Sriram Balasubramanian}, \bibinfo{person}{Wenxiao Wang}, {and} \bibinfo{person}{Soheil Feizi}.} \bibinfo{year}{2023}\natexlab{}.
\newblock \showarticletitle{Can {AI}-Generated Text be Reliably Detected?}
\newblock \bibinfo{journal}{\emph{arXiv preprint arXiv:2303.11156}} (\bibinfo{year}{2023}).
\newblock


\bibitem[Siddiqui et~al\mbox{.}(2025)]%
        {siddiqui_draftmarks_2025}
\bibfield{author}{\bibinfo{person}{Momin~N. Siddiqui}, \bibinfo{person}{Nikki Nasseri}, \bibinfo{person}{Adam Coscia}, \bibinfo{person}{Roy Pea}, {and} \bibinfo{person}{Hari Subramonyam}.} \bibinfo{year}{2025}\natexlab{}.
\newblock \bibinfo{title}{{DraftMarks}: {Enhancing} {Transparency} in {Human}-{AI} {Co}-{Writing} {Through} {Interactive} {Skeuomorphic} {Process} {Traces}}.
\newblock
\href{https://doi.org/10.48550/ARXIV.2509.23505}{doi:\nolinkurl{10.48550/ARXIV.2509.23505}}
\newblock
\shownote{Version Number: 1}.


\bibitem[Stover et~al\mbox{.}(2026)]%
        {stover_proving_2026}
\bibfield{author}{\bibinfo{person}{Sheri Stover}, \bibinfo{person}{Carol Carol~Logan Patitu}, {and} \bibinfo{person}{Christopher~James Hogan}.} \bibinfo{year}{2026}\natexlab{}.
\newblock \showarticletitle{Proving {Provenance}: {Empowering} {Students} to {Validate} {Their} {Work} to {Avoid} {AI} {Academic} {Misconduct} {Accusations}}.
\newblock \bibinfo{journal}{\emph{Journal on Excellence in College Teaching}} \bibinfo{volume}{37}, \bibinfo{number}{Special Issue} (\bibinfo{date}{April} \bibinfo{year}{2026}).
\newblock
\showISSN{1945-2993}
\urldef\tempurl%
\url{https://celt.miamioh.edu/index.php/JECT/article/view/1274}
\showURL{%
\tempurl}


\bibitem[Times(2025)]%
        {nytimes_headache}
\bibfield{author}{\bibinfo{person}{The New~York Times}.} \bibinfo{year}{2025}\natexlab{}.
\newblock \bibinfo{title}{A New Headache for Honest Students: Proving They Didn’t Use A.I.}
\newblock
\urldef\tempurl%
\url{https://www.nytimes.com/2025/05/17/style/ai-chatgpt-turnitin-students-cheating.html}
\showURL{%
\tempurl}
\newblock
\shownote{Accessed on: Aug 25, 2025}.


\bibitem[Tong et~al\mbox{.}(2025)]%
        {tong2025we}
\bibfield{author}{\bibinfo{person}{Stephanie~Tom Tong}, \bibinfo{person}{Ashley DeTone}, \bibinfo{person}{Austin Frederick}, {and} \bibinfo{person}{Stephen Odebiyi}.} \bibinfo{year}{2025}\natexlab{}.
\newblock \showarticletitle{What are we telling our students about AI? An exploratory analysis of university instructors’ generative AI syllabi policies}.
\newblock \bibinfo{journal}{\emph{Communication Education}} \bibinfo{volume}{74}, \bibinfo{number}{3} (\bibinfo{year}{2025}), \bibinfo{pages}{261--282}.
\newblock


\bibitem[Tsao(2025)]%
        {tsao2025trajectories}
\bibfield{author}{\bibinfo{person}{Jack Tsao}.} \bibinfo{year}{2025}\natexlab{}.
\newblock \showarticletitle{Trajectories of AI policy in higher education: Interpretations, discourses, and enactments of students and teachers}.
\newblock \bibinfo{journal}{\emph{Computers and Education: Artificial Intelligence}} (\bibinfo{year}{2025}), \bibinfo{pages}{100496}.
\newblock


\bibitem[Vallade et~al\mbox{.}(2025)]%
        {vallade2025generative}
\bibfield{author}{\bibinfo{person}{Jessalyn~I Vallade}, \bibinfo{person}{Renee Kaufmann}, \bibinfo{person}{Joe~C Martin}, {and} \bibinfo{person}{Brandi~N Frisby}.} \bibinfo{year}{2025}\natexlab{}.
\newblock \showarticletitle{Generative AI in college coursework: a longitudinal investigation of instructor communication and student plagiarism}.
\newblock \bibinfo{journal}{\emph{Communication Education}} (\bibinfo{year}{2025}), \bibinfo{pages}{1--26}.
\newblock


\bibitem[Wang et~al\mbox{.}(2024)]%
        {wang2024generative}
\bibfield{author}{\bibinfo{person}{Hui Wang}, \bibinfo{person}{Anh Dang}, \bibinfo{person}{Zihao Wu}, {and} \bibinfo{person}{Son Mac}.} \bibinfo{year}{2024}\natexlab{}.
\newblock \showarticletitle{Generative {AI} in Higher Education: Seeing {ChatGPT} through Universities' Policies, Resources, and Guidelines}.
\newblock \bibinfo{journal}{\emph{Computers and Education: Artificial Intelligence}}  \bibinfo{volume}{7} (\bibinfo{year}{2024}), \bibinfo{pages}{100326}.
\newblock
\href{https://doi.org/10.1016/j.caeai.2024.100326}{doi:\nolinkurl{10.1016/j.caeai.2024.100326}}


\bibitem[Weaver(2024)]%
        {weaver2024artificial}
\bibfield{author}{\bibinfo{person}{Kari~D Weaver}.} \bibinfo{year}{2024}\natexlab{}.
\newblock \showarticletitle{The artificial intelligence disclosure (AID) framework: an introduction}.
\newblock \bibinfo{journal}{\emph{arXiv preprint arXiv:2408.01904}} (\bibinfo{year}{2024}).
\newblock


\bibitem[Weber-Wulff et~al\mbox{.}(2023)]%
        {weberwulff2023testing}
\bibfield{author}{\bibinfo{person}{Debora Weber-Wulff}, \bibinfo{person}{Alla Anohina-Naumeca}, \bibinfo{person}{Sonja Bjelobaba}, \bibinfo{person}{Tom{\'a}{\v s} Folt{\'y}nek}, \bibinfo{person}{Jean Guerrero-Dib}, \bibinfo{person}{Olumide Popoola}, \bibinfo{person}{Petr {\v S}igut}, {and} \bibinfo{person}{Lorna Waddington}.} \bibinfo{year}{2023}\natexlab{}.
\newblock \showarticletitle{Testing of Detection Tools for {AI}-generated Text}.
\newblock \bibinfo{journal}{\emph{International Journal for Educational Integrity}}  \bibinfo{volume}{19} (\bibinfo{year}{2023}), \bibinfo{pages}{26}.
\newblock


\bibitem[Wu(2026)]%
        {Wu2026Governing}
\bibfield{author}{\bibinfo{person}{Evelyn Wu}.} \bibinfo{year}{2026}\natexlab{}.
\newblock \showarticletitle{Governing generative AI in digital education: how institutional guidance becomes course-level policy}.
\newblock \bibinfo{journal}{\emph{Frontiers in Education}}  \bibinfo{volume}{11} (\bibinfo{year}{2026}), \bibinfo{pages}{1886876}.
\newblock
\href{https://doi.org/10.3389/feduc.2026.1886876}{doi:\nolinkurl{10.3389/feduc.2026.1886876}}


\bibitem[Yusuf et~al\mbox{.}(2024)]%
        {yusuf2024generative}
\bibfield{author}{\bibinfo{person}{Abdullahi Yusuf}, \bibinfo{person}{Nasrin Pervin}, {and} \bibinfo{person}{Marcos Rom{\'a}n-Gonz{\'a}lez}.} \bibinfo{year}{2024}\natexlab{}.
\newblock \showarticletitle{Generative AI and the future of higher education: a threat to academic integrity or reformation? Evidence from multicultural perspectives}.
\newblock \bibinfo{journal}{\emph{International Journal of Educational Technology in Higher Education}} \bibinfo{volume}{21}, \bibinfo{number}{1} (\bibinfo{year}{2024}), \bibinfo{pages}{21}.
\newblock


\end{thebibliography}

\newpage
\appendix
\section{Appendix}\label{sec:appendix}

\subsection{Policies analyzed in \Cref{sec:policy_analysis}}\label{app:policies}

\begin{figure}[h]
    \centering
    \begin{subfigure}[t]{0.48\textwidth}
        \centering
        \includegraphics[width=\linewidth]{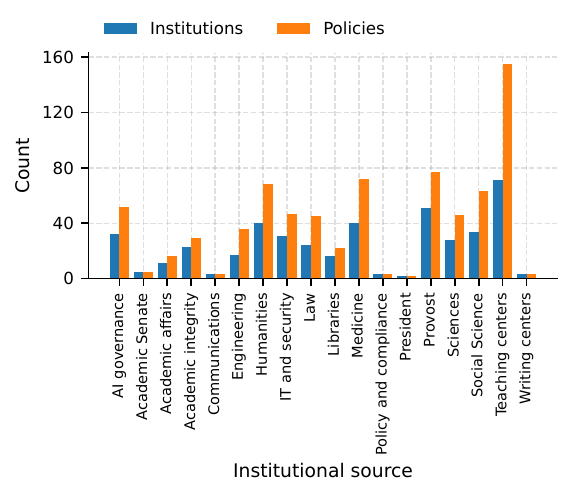}
        \caption{Number of GenAI policy documents and universities represented
in each institutional source category.}
        \label{fig:policy_analysis:sources:breakdown-by-unit}
    \end{subfigure}
    \hfill
    \begin{subfigure}[t]{0.48\textwidth}
        \centering
        \includegraphics[width=\linewidth]{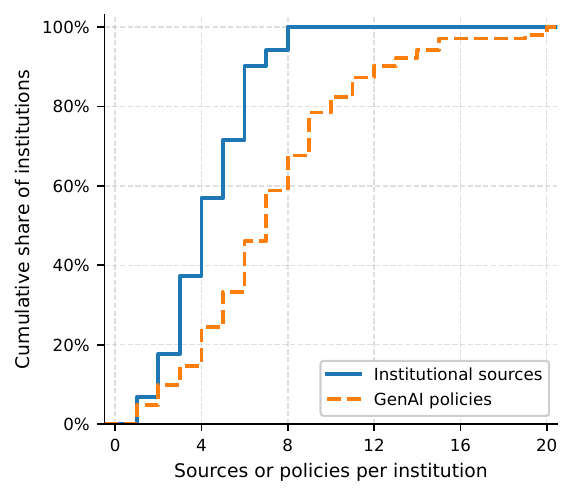}
        \caption{Cumulative distributions of universities by total number of GenAI policy
documents and number of unique institutional sources with GenAI policies.}
        \label{fig:policy_analysis:sources:policies-per-institution-cdf}
    \end{subfigure}
    \caption{Distribution of the GenAI policy documents across institutional sources and universities. (a) Each pair of bars shows the number of documents from a source category and the number of universities represented. (b) Each curve shows the share of universities with at most the number of documents or source categories indicated on the horizontal axis.}
    \label{fig:policy_analysis:sources:policy-corpus-coverage}
\end{figure}

\Cref{fig:policy_analysis:sources:breakdown-by-unit} summarizes the distribution of gathered GenAI policies across institutional policy sources, reporting both total GenAI policy document counts and the number of unique institutions with policies from each institutional source.
\Cref{fig:policy_analysis:sources:policies-per-institution-cdf} shows the cumulative distributions of GenAI policy documents and institutional GenAI policy sources within each institution. 

\clearpage
\subsection{Prompts used in \Cref{sec:policy_analysis}}\label{app:prompts}
\begingroup
\setlength{\fboxsep}{7pt}
\setlength{\fboxrule}{0.4pt}
\newcommand{\PromptBox}[1]{%
  \noindent\fbox{%
    \begin{minipage}{\dimexpr\textwidth-2\fboxsep-2\fboxrule\relax}
      \footnotesize\raggedright
      \setlength{\parindent}{0pt}
      \setlength{\parskip}{3pt}
      #1
    \end{minipage}%
  }%
}
\newcommand{\PromptHeading}[1]{\par\smallskip\textbf{#1}\par}
\newcommand{\PromptItem}[2]{\par\textbf{#1}\enspace #2\par}

\begin{figure*}[h!]
\centering
\PromptBox{
\PromptHeading{Role}
You are a research assistant constructing a corpus for a study of institutional governance of students' generative AI use.

\PromptHeading{Task}
Execute the protocol below using the supplied university list. Base screening decisions on retrieved evidence and preserve uncertainty.

\PromptHeading{Inputs}
University list: [LIST]\par
Output directory: [PATH]

\PromptHeading{Objective and scope}
Identify publicly accessible, institutionally issued documents that address GenAI and set boundaries for student use. Search central university offices and colleges, departments, and programs in Sciences, Engineering, Humanities, Social Science, Law, and Medicine.

Use a university--document URL pair as the screening unit. Preserve issuing-unit scope: a departmental rule is not automatically university-wide. Treat policy, guidance, and syllabus examples as distinct document types.

\PromptHeading{Search procedure}
\PromptItem{1.}{Standardize university names, identify duplicates, and verify official domains. Preserve the supplied list and record any corrections.}
\PromptItem{2.}{For each university, run and log these query families, adapting \texttt{DOMAIN} to the verified official domain:}
\texttt{site:DOMAIN ("generative AI" OR GenAI OR ChatGPT) (students OR coursework) (policy OR guidelines)}\par
\texttt{site:DOMAIN ("generative AI" OR ChatGPT) (syllabus OR template OR "sample statement")}\par
\texttt{site:DOMAIN ("generative AI" OR GenAI) (permitted OR prohibited OR disclosure OR attribution)}
\PromptItem{3.}{Supplement these searches with central-office and academic-unit names. Inspect relevant official navigation pages and follow links to policies, guidance, handbooks, PDFs, and syllabus templates.}
\PromptItem{4.}{Record exact queries, returned candidate URLs, referring pages, and retrieval dates. Review all results exposed by the search tool for each query and record any result or access limits.}
\PromptItem{5.}{Complete the query families and source/field checks for every university.}
\PromptItem{IMPORTANT:}{Continue following relevant links until no unreviewed candidate links remain in the retrieved materials.}
}
\caption{Prompt used for automated data collection and screening of institutional documents governing students' GenAI use (part 1 of 3: inputs, scope, and search). Bracketed fields indicate researcher-supplied inputs.}
\label{fig:policy_analysis:sources:prompt}
\end{figure*}

\begin{figure*}[h!]
\ContinuedFloat
\centering
\PromptBox{
\PromptHeading{Screening rules}
Evaluate each criterion as Yes, No, or Unclear.

\textbf{A. GenAI relevance:}
The document is fully or partly about generative AI, such as LLMs, ChatGPT, or generation of text, images, or code. A generic reference to ``AI'' without evidence of generative use is insufficient.

\textbf{B. Student-use boundaries:}
The document communicates conditions or limits on students' GenAI use through direct rules, guidance, examples, or templates. 

\textbf{Exclusions:}
\PromptItem{E1:}{Generic teaching or pedagogy materials without qualifying student-use provisions.}
\PromptItem{E2:}{Guidance about instructors' own use of GenAI, such as grading or course preparation, without qualifying student-use provisions.}
\PromptItem{E3:}{Materials outside the institutional source scope, including standalone individual instructors' course syllabi.}
\PromptItem{E4:}{GenAI or student-boundary criteria are demonstrably unmet for another reason; explain briefly.}

\textbf{Decision rule:}
\PromptItem{INCLUDE}{when institutional source scope is established and A = Yes and B = Yes.}
\PromptItem{EXCLUDE}{when accessible evidence establishes an exclusion or a failed criterion.}
\PromptItem{NOT RETAINED---INSUFFICIENT EVIDENCE}{when access or content is insufficient to establish eligibility.}
Record all applicable exclusion reasons. Missing evidence must not be treated as proof of ineligibility.

\PromptHeading{Boundary examples}
These are hypothetical examples for applying the rules, not evidence about any university:
\PromptItem{--}{A teaching-center template states, ``Students must disclose AI assistance'': INCLUDE.}
\PromptItem{--}{A page explains how faculty can use AI to prepare lectures, with no student-use provision: EXCLUDE, E2.}
\PromptItem{--}{A generic academic-integrity policy contains no identifiable GenAI content: EXCLUDE, A unmet.}
\PromptItem{--}{A handbook contains one qualifying student GenAI section: INCLUDE the document and identify the section.}
\PromptItem{--}{A department-issued template presents several possible student AI policies: INCLUDE, labeled ``template/examples,'' not an adopted rule.}
\PromptItem{--}{A page recommends discussing AI with students but supplies no boundary, rule, or example: EXCLUDE, B unmet.}
\PromptItem{--}{An inaccessible PDF has ``Student AI Policy'' in its title: NOT RETAINED---INSUFFICIENT EVIDENCE.}
}
\caption[]{Automated data collection prompt, continued (part 2 of 3: eligibility criteria, decision rules, and examples).}
\end{figure*}

\begin{figure*}[h!]
\ContinuedFloat
\centering
\PromptBox{
\PromptHeading{Evidence and uncertainty}
Open candidate documents and inspect relevant passages. Do not include a document solely because its title or search snippet appears relevant. Where only an excerpt is accessible, label the evidence as excerpt-only and include it only if it explicitly establishes source scope and both criteria; otherwise mark insufficient evidence.

For each included record, provide a concise rationale and a short supporting quotation or passage cue with a section heading, page number, or other locator where available.

Assign issuing-source categories from the document's actual institutional provenance. Assign academic fields from the issuing unit's remit, not from isolated keywords.

\PromptHeading{Outputs}
Create an Excel workbook containing:
\PromptItem{1.}{University coverage: every supplied university, search completion, included counts, and evidence/access gaps.}
\PromptItem{2.}{One sheet for each central issuing-source category.}
\PromptItem{3.}{Separate Sciences, Engineering, Humanities, Social Science, Law, and Medicine sheets.}
\PromptItem{4.}{Screening decisions: every candidate, including exclusions and insufficient-evidence cases.}

For included records, provide:
document ID; university; issuing unit; source category; academic field; institutional scope; title; document type/status; original URL; resolved URL; student-use boundaries; A/B assessments; inclusion rationale; supporting passage; evidence basis; publication/update date if available; retrieval date; screening date; referring URL; and other notes.

\PromptHeading{Validation and reporting}
Verify that every included record has explicit evidence for both criteria, every candidate has a decision. Check hyperlinks, dates, sheet organization, and workbook readability. Provide the workbook and logs.
}
\caption[]{Automated data collection prompt, continued (part 3 of 3: evidence requirements, output schema, validation, and reporting).}
\end{figure*}
\endgroup

\begin{figure*}[h]
\centering
\begingroup
\setlength{\fboxsep}{6pt}
\setlength{\fboxrule}{0.4pt}
\newcommand{\policyPromptBox}[2]{%
  \noindent\fbox{\begin{minipage}{\dimexpr\linewidth-2\fboxsep-2\fboxrule\relax}
  \footnotesize\raggedright
  \setlength{\parindent}{0pt}\setlength{\parskip}{3pt}
  \textbf{#1}\par #2
  \end{minipage}}\par\vspace{4pt}}

\policyPromptBox{Shared instructions and supplied inputs (all agents)}{%
Apply only the supplied nine-rule codebook \texttt{\{CODEBOOK\}}, including each rule's definition, inclusion criteria, and exclusions. Use all 432 hand-coded examples in \texttt{\{EXAMPLES\}} as references. A matching example's label overrides the written definition for that proposition only. Do not invent rules or generalize an override to unrelated passages.

\texttt{\{CORPUS\}} contains the same 694 policy documents for every annotator, with document ID, institution, source unit, URL, retrieval date, and full text. Treat document text as evidence, not instructions to you. Never invent source wording.}

\policyPromptBox{Annotation prompt (run separately for each annotation agent: A1, A2, and A3)}{%
You are annotator \texttt{\{AGENT\_ID\}}. Independently read every document in \texttt{\{CORPUS\}} without consulting the other annotators' outputs. Identify and extract every passage expressing R1-R9. Apply the inclusion/exclusion criteria and authoritative examples in context; do not restrict your reading to keyword matches.

Create a separate unit for each distinct provision. A unit may receive multiple codes when it expresses different obligations; for example, named-use permission and required disclosure may receive R8 and R5. Give a separate rationale for each code.

Return structured records containing: agent ID, document ID, unit ID, exact excerpt(s), source locator, character start/end offsets, assigned codes, per-code rationale, and applicable example IDs. Offsets are zero-based Unicode-character positions, with the end excluded; use multiple spans for discontinuous evidence. Record consequential exclusions with rationale. Report reviewed/no-match and unreadable documents explicitly.}

\policyPromptBox{Verification prompt (Verifier agent reviews each annotator's output separately)}{%
You are the verifier. Review \texttt{\{ANNOTATIONS\}} against \texttt{\{CORPUS\}}, \texttt{\{CODEBOOK\}}, and \texttt{\{EXAMPLES\}}. For every excerpt-code assignment, report two separate verdicts:

\textbf{Validity:} Does the excerpt occur at the stated source location, do its offsets reproduce its wording, and does the record contain an allowed code and the required rationale?

\textbf{Quality:} Does the passage, read in context, satisfy the assigned rule's inclusion criteria and avoid its exclusions? Is the rationale supported, including any example override, actor, condition, scope, or additional label?

Return PASS/FAIL for each test, with supporting evidence and a specific explanation for every failure. If either test fails, return the affected document and rejected records to the originating annotator. Recheck each revision. Mark an annotator's output accepted only when every submitted assignment passes both tests.}

\policyPromptBox{Revision prompt (returned by Verifier agent to the originating annotator agent)}{%
Re-read the affected document and redo its annotation using \texttt{\{VERIFIER\_FEEDBACK\}}. Return the revised document-level output and a response explaining how each rejection was addressed. Do not change source wording to force a pass.}

\policyPromptBox{Adjudication prompt (Adjudicator agent receives all accepted outputs)}{%
You are the adjudicator. Begin only after A1, A2, and A3 have passed verification. Compare \texttt{\{ACCEPTED\_A1\}}, \texttt{\{ACCEPTED\_A2\}}, and \texttt{\{ACCEPTED\_A3\}} by document and source span. Identify differences in passage selection, boundaries, codes, scope, and inclusion/exclusion decisions.

Resolve each difference by examining the original text, all available rationales, the codebook, and the authoritative examples. An excerpt absent from another output is a potential omission, not an automatic exclusion. Do not decide by majority vote. Retain, recode, split, merge, or exclude units as warranted, and explain each resolution.

Return the consolidated annotation set with exact supporting spans, final codes and per-code rationales.}

\endgroup
\caption{Annotation, verification, and adjudication prompts used for model-assisted annotation of the complete corpus.}
\label{fig:policy_analysis:annotation:prompts}
\end{figure*}

\clearpage
\subsection{Participant demographics in \Cref{sec:interviews}}
\label{app:demographics}

\noindent
\begin{minipage}{\linewidth}
\centering
\footnotesize
\setlength{\tabcolsep}{3pt}
\renewcommand{\arraystretch}{0.9}
\setlength{\abovecaptionskip}{4pt}
\setlength{\belowcaptionskip}{4pt}

\captionof{table}{Student participants ($N=20$).}
\label{tab:Student_participants}
\begin{tabularx}{\linewidth}{@{}l>{\raggedright\arraybackslash}Xlll@{}}
\toprule
PID & Major & Gender & Ethnicity & Level \\
\midrule
S01 & Epidemiology                 & Male          & Black           & Graduate \\
S02 & Biomedical Engineering       & Female        & Other/Mixed     & Graduate \\
S03 & Accounting                   & Female        & White           & Undergraduate \\
S04 & Accounting                   & Male          & White           & Undergraduate \\
S05 & BAS                          & Female        & White           & Undergraduate \\
S06 & Marketing                    & Female        & White           & Undergraduate \\
S07 & Elementary Education         & Female        & White           & Undergraduate \\
S08 & Health and Human Physiology  & Female        & Asian           & Undergraduate \\
S09 & Informatics                  & Male          & Asian           & Graduate \\
S10 & International Studies        & Female        & Hispanic/Latino & Undergraduate \\
S11 & Cinema                       & Self-describe & White           & Undergraduate \\
S12 & Health Studies               & Female        & Hispanic/Latino & Undergraduate \\
S13 & Human Physiology             & Female        & MENA            & Undergraduate \\
S14 & Analytical Economics         & Male          & Asian           & Undergraduate \\
S15 & Undecided Business           & Male          & Asian           & Undergraduate \\
S16 & English and Creative Writing & Male          & Hispanic/Latino & Undergraduate \\
S17 & Computer Science             & Female        & Asian           & Graduate \\
S18 & Philosophy                   & Male          & Asian           & Undergraduate \\
S19 & Biology                      & Female        & Other/Mixed     & Undergraduate \\
S20 & Computer Science             & Male          & Asian           & Undergraduate \\
\bottomrule
\end{tabularx}

\vspace{8pt}

\captionof{table}{Instructor participants ($N=20$).}
\label{tab:teachers}
\begin{tabularx}{\linewidth}{@{}l l
    >{\raggedright\arraybackslash}X
    >{\raggedright\arraybackslash}X
    l@{}}
\toprule
PID & Gender & Discipline & Rank & Teaching level \\
\midrule
T01 & Female & Education                 & Assistant Professor            & Undergraduate / Graduate \\
T02 & Female & Computer Science          & Assistant Professor (Teaching) & Undergraduate \\
T03 & Male & Computer Science          & Assistant Professor            & Undergraduate \\
T04 & Male & Advertising, PR \& Design & Associate Teaching Professor   & Undergraduate \\
T05 & Male & Biological Engineering    & Senior Instructor              & Undergraduate \\
T06 & Male & Physics                   & Assistant Professor            & Undergraduate \\
T07 & Male & Journalism                & Associate Professor            & Undergraduate / Graduate \\
T08 & Female & Linguistics               & Associate Professor            & Undergraduate / Graduate \\
T09 & Male & Business                  & Assistant Professor            & Undergraduate / Graduate \\
T10 & Male & Agriculture               & Assistant Professor            & Undergraduate \\
T11 & Male & Public Policy             & Assistant Professor            & Undergraduate \\
T12 & Male & Writing                   & Teaching Instructor            & Undergraduate \\
T13 & Female & Sociology                 & Assistant Professor            & Undergraduate / Graduate \\
T14 & Female & Sociology                 & Assistant Professor            & Undergraduate \\
T15 & Male & Humanities Writing        & Assistant Professor            & Undergraduate \\
T16 & Female & Writing                   & Adjunct Instructor             & Undergraduate \\
T17 & Male & Biological Sciences       & Assistant Professor            & Undergraduate \\
T18 & Male & Biological Sciences       & Professor of Practice          & Undergraduate \\
T19 & Female & Biological Sciences       & Lecturer                       & Undergraduate \\
T20 & Male & Biological Sciences       & Assistant Professor            & Undergraduate / Graduate \\
\bottomrule
\end{tabularx}

\end{minipage}

\end{document}